\documentclass[aps,prb,preprint,groupedaddress]{revtex4-1}
\usepackage{graphicx}
\begin{document}

\title{Semiconductor-metal transition in semiconducting bilayer sheets of transition metal dichalcogenides} 
\author{Swastibrata Bhattacharyya}
\affiliation {Materials Research Centre, Indian Institute of Science, Bangalore
  560012, India}
\author{Abhishek K. Singh}
\email {abhishek@mrc.iisc.ernet.in}
\affiliation {Materials Research Centre, Indian Institute of Science, Bangalore
  560012, India}

\date{\today}
\begin{abstract}
Using first-principles calculations we show that the band gap of bilayer sheets of semiconducting transition metal dichalcogenides (TMDs) can be reduced smoothly by applying vertical compressive pressure. These materials undergo a universal reversible semiconductor to metal (S-M) transition at a critical pressure. S-M transition is attributed to lifting the degeneracy of the bands at fermi level caused by inter-layer interactions via charge transfer from metal to chalcogens. The S-M transition can be reproduced even after incorporating the band gap corrections using hybrid functionals and GW method. The ability to tune the band gap of TMDs in a controlled fashion over a wide range of energy, opens-up possibility for its usage in a range of applications.
\end{abstract}

\maketitle 
\section{INTRODUCTION}
Despite being the most promising two-dimensional (2D) material gapless graphene has limitations for its applications in nanoelectronics and nanophotonics. This led to the finding of other 2D-materials with finite band gap such as BN, transition metal dichalcogenides (TMDs) and transition metal oxides (MO$_2$). BN sheet is an insulator and modifying their band gap for optical and electronic applications is still a challenge. Depending on the combination of metal and chalcogens, TMDs offer a wide range of 2D materials: metals \cite {DingMX22011,ayari2007}, superconductors \cite{Gabovich2001, Morosan2006}, charge density wave systems \cite {Rossnagel2001, Hu2007}, Mott insulators\cite {Kusmartseva2008} and semiconductors \cite{KamBGap1982, Han2011}. Semiconducting two-dimensional TMDs include  MoS$_2$,  MoSe$_{2}$,  MoTe$_2$,  WS$_2$ and  WSe$_2$ and have emerged as promising materials for a range of applications.

Few layers to monolayer of  MoS$_2$ and other TMDs have been successfully synthesized \cite {Frindt1963, Novoselov2D2005, frindtMoS1966, Joensen1986, Schumacher1993, ColemanLiquid2011, miremadiWS1988, YangWS21996, Ramak2010, Eda2011} and their optical absorption and photoconductivity have been studied \cite {Frindt1963, korn2011}. A single layer of TMDs having stoichiometry of MX$_2$, consists of a hexagonally arranged transition metal (M = Ti, Nb, Ta, Mo, and W) sandwiched between two layers of chalcogen atoms (X = S, Se, and Te). Within a layer the metals and chalcogens form strong ionic-covalent bonds, whereas in bulk TMDs, the layers are bonded by weaker van der Waals (vdW) interaction. The bulk TMDs are indirect band gap semiconductors having band gaps in the range of 1.0-1.35 eV \cite{KamBGap1982}. With the reduction in number of layers the band gap increases \cite{Han2011} and transforms into a direct gap for a single layer TMD \cite {Lebegue2009, Splendiani2010,  ellis2011}. In order to use  these materials as building blocks in nanoelectronics, their electronic properties need to be modified. This has been achieved by doping\cite {DingMX22011, Inosov2008} and intercalation \cite {Kidd2011, Colev2009}.  The TMD based field effect transistors with high room-temperature current on/off ratios \cite {RadisavljevicNatnano} and higher on- current density \cite {Liu2011, Yoon2011} as well as integrated circuits \cite {Radisavljevic2011} have been successfully fabricated.

Tuning band gap of 2D materials---for their potential application in electromechanical devices, tunable photodetectors and lasers---has been a challenge in band gap engineering. Applied strain or electric field offers a novel way of modifying the band gaps over a wide range.
Even for graphene, it has been shown using density functional theory calculations that uniaxial strain on monolayers \cite {Zhen2008} and applied vertical electric field to bilayers \cite {McCann2006, Castro2007, Ohta2006, Zhang2009, Xia2010} open a small band gap. Similar, theoretical studies for TMDs show the semiconductor to metal (S-M) transition for bilayers subjected to vertical electric field (0.2-0.3 V/\AA)\cite {Ramasubramaniam2011} and for mono- and bilayers under biaxial strain\cite {Scalise2012}. These techniques are promising but suffer from the practical applicability. For example, the electric field required to achieve S-M transition is too large, and a reversible way of applying a biaxial strain in 2D materials has yet to be demonstrated in laboratory. On the other hand it has been experimentally shown that the band structure of bulk TMDs can be modified by application of compressive strain \cite {Dave2004}. Here, we investigate the effect of normal compressive strain (NCS) on electronic properties of semiconducting bilayer TMDs:  MoS$_2$,  MoSe$_2$,  MoTe$_2$,  WS$_2$ and  WSe$_2$. The band gaps of these materials decrease gradually with the applied NCS.  A reversible semiconductor to metal  transition was observed after a threshold pressure P$_{th}$ was achieved. The P$_{th}$ depends upon the material as well as the stacking pattern of the two layers. The hybrid functional and GW methods were used to correct the PBE gap. The S-M transition was found to be independent of the methods. This offers a wide range (~1.9 - 0.0 eV) of reversible band gap tuning, which can be utilized for various applications.

\begin{figure}[t]
\centering
\includegraphics[width=\columnwidth]{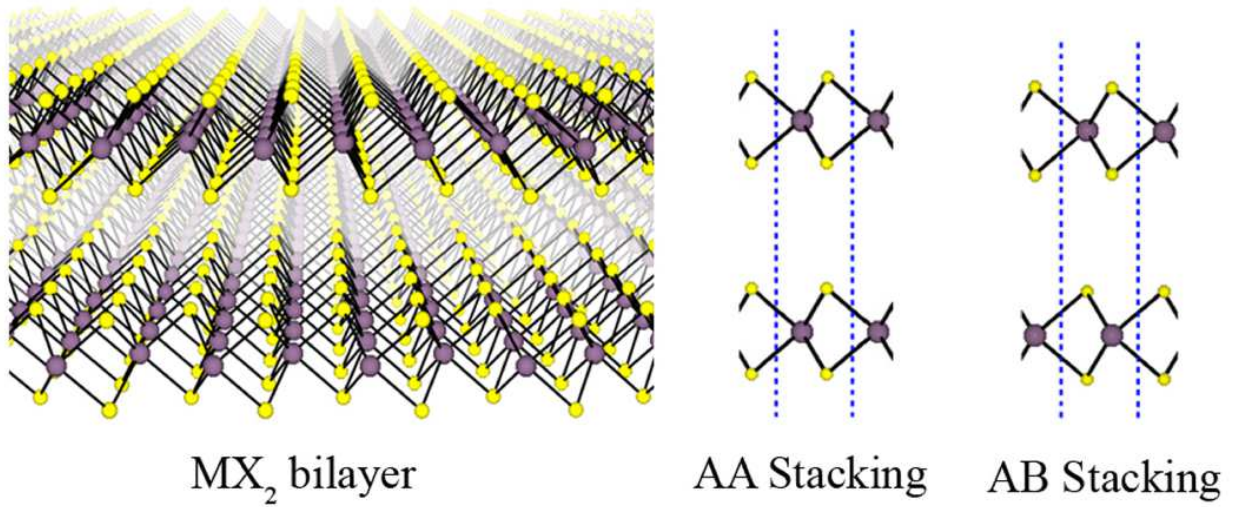}
\caption{ (Color online) Structure of AB-stacked MX$_2$ bilayer with M and X atoms are shown by purple and yellow spheres, respectively. Side view of the bilayers with AA and AB stacking. The blue dotted lines show the unit cells.}
\label{fig:1}
\end{figure}

\section{Computational Details}

\begin{table}[b]
\centering
\begin{tabular}{c c c c c c c c c c}
\hline
\hline
\multicolumn{1}{c}{ } & \multicolumn{1}{c}{  } &  \multicolumn{4}{c}{ Structural Parameters  (\AA)} &
 \multicolumn{4}{c} {Band Gap (eV) } \\
\multicolumn{1}{c}{Material } & \multicolumn{1}{c}{ Functional } & \multicolumn{1}{c}{ a } & \multicolumn{1}{c}{ c } &
\multicolumn{2}{c}{Bilayer separation } &  \multicolumn{1}{c}{Bulk} & \multicolumn{1}{c}{Monolayer}  &  \multicolumn{2}{c}{ Bilayer}\\

\multicolumn{1}{c}{} & \multicolumn{1}{c}{} & \multicolumn{1}{c}{ } & \multicolumn{1}{c}{  } &
\multicolumn{1}{c}{AB} & \multicolumn{1}{c}{AA} & \multicolumn{1}{c}{} & \multicolumn{1}{c}{  } & \multicolumn{1}{c}{ AB} & \multicolumn{1}{c}{ AA} \\

\hline

                        &  PBE& 3.18 & 12.42$^{**}$ & 4.14 & 4.54 & 0.94 & 1.68 & 1.59 & 1.63  \\
  MoS$_2$      & PBE + vdW & 3.21& 12.42 & 3.11 & 3.72 & 0.85 & \--  & 1.25 & 1.47  \\    
                        & Exp. & 3.16 \cite{Boker2001} & 12.29\cite{Boker2001}  & \--- & \-- &  1.23\cite {KamBGap1982} & 1.9\cite {Mak2010} & 1.6\cite {Mak2010} & \--   \\
\hline                     
                        & PBE & 3.32 & 13.06$^{**}$ & 3.88 & 4.36 &  0.87 & 1.43 & 1.42  & 1.41  \\
 MoSe$_2$	   & PBE + vdW & 3.33  &13.06         & 3.19 &3.78 & 0.87   & \-- & 1.18  & 1.41 \\
                        & Exp. & 3.30\cite{Boker2001} & 12.94\cite{Boker2001}      & \--     &\--    &  1.09\cite {KamBGap1982} & \--  &\--   & \--  \\
\hline                         
                        & PBE & 3.56 &13.95$^{**}$& 3.95 & 4.97 &  0.67 & 1.06 & 1.03  & 1.05  \\                       
 MoTe$_2$	   & PBE + vdW & 3.54 & 13.95 & 3.37 & 4.07 & 0.68  & \-- & 0.88 & 1.02 \\
                        & Exp. & 3.52\cite{Boker2001} & 13.97\cite{Boker2001} & \-- & \-- & 1.00\cite{Boker2001}  &\--  &  \-- &  \-- \\
\hline                         
                        & PBE & 3.18 & 12.99$^{**}$& 4.31 &  4.56 & 1.25  & 1.81 & 1.77  &  1.79 \\                    
 WS$_2$ 	   & PBE + vdW &3.19 & 12.99 & 3.39 & 3.95  &  1.24 & \-- & 1.57 & 1.70  \\
                        & Exp. &3.15\cite{Schutte1987} & 12.32\cite{Schutte1987} & \-- & \--& 1.35\cite {KamBGap1982}  &  \--& \--  & \--  \\
\hline                         
                        & PBE & 3.32 & 13.38$^{**}$&4.11  & 4.43 & 1.09  & 1.53 & 1.51 & 1.52 \\
 WSe$_2$	   & PBE + vdW & 3.34  & 13.38 & 3.35  & 3.98 & 1.08  & \--  & 1.43 & 1.51  \\
                        & Exp. & 3.28\cite{Schutte1987} & 12.96\cite{Schutte1987} & \-- & \--  & 1.20\cite {KamBGap1982}  &\--  & \-- & \--  \\

\hline
\hline
\end{tabular}
\\ $^{**}$Same values as vdW 
\caption{Structural parameters and band gaps of bulk, monolayer and bilayer of TMDs calculated using PBE, PBE+vdW. For comparison purpose the corresponding experimental values are also included.}
\label{table:1}
\end{table}

The calculations were performed using ab-initio density functional theory (DFT) in conjunction with all-electron projector augmented wave potentials \cite {Blochl94, Kresse99}  and the Perdew-Burke-Ernzerhof \cite {Perdew96} generalized gradient approximation (GGA) to the electronic exchange and correlation, as implemented in the Vienna Ab initio Simulation Package (VASP)\cite {Kresse1993}.  We optimize the structure of bilayer TMDs (MX$_2$ with M = Mo, W and X = S, Se, Te), using the unit cells as shown in  Fig.~\ref{fig:1}.  A well converged Monkhorst-Pack k-point set ($15 \times 15 \times 1$) was used for the calculation and conjugate gradient scheme was employed to optimize the geometries until the forces on every atom were $\leq 0.005$ eV/\AA. Sufficient vacuum was used along z direction i.e. perpendicular to the 2D sheet, to avoid spurious interaction among the periodic images. The lattice parameter and interlayer distance of the optimized structures are listed in Table~\ref{table:1} and are in good agreement with the previously reported PBE values \cite {DingMX22011}. 

The week van der Waals interaction between the layers has an effect in determining the interlayer distance for the bilayers as well as for the bulk MX$_2$. The van der Waals interaction originates from dynamical correlations between fluctuating charge distributions and cannot be described by the PBE functional. Consequently,  the relaxed bilayer distance obtained by PBE is off by  approximately 1~\AA~compared to the bulk interlayer distance. We incorporate the van der Waals interactions through a pair-wise force field following Grimme's DFT-D2 method\cite {Grimme2006}. The parameters used in DFT-D2 method are thoroughly optimized for several of the DFT functionals, including PBE. We performed a comprehensive test on the reliability of the empirical parameter by calculating the bulk phase of TMDs. The lattice parameters obtained using this approach are in very good agreement with the experimental values \cite{Boker2001,Schutte1987}, Table~\ref{table:1}. The calculated DFT-D2 interlayer distances in bilayer TMDs have decreased significantly compared with PBE, Table~\ref{table:1}.

\begin{figure}[t]
\centering
\includegraphics[width=\columnwidth]{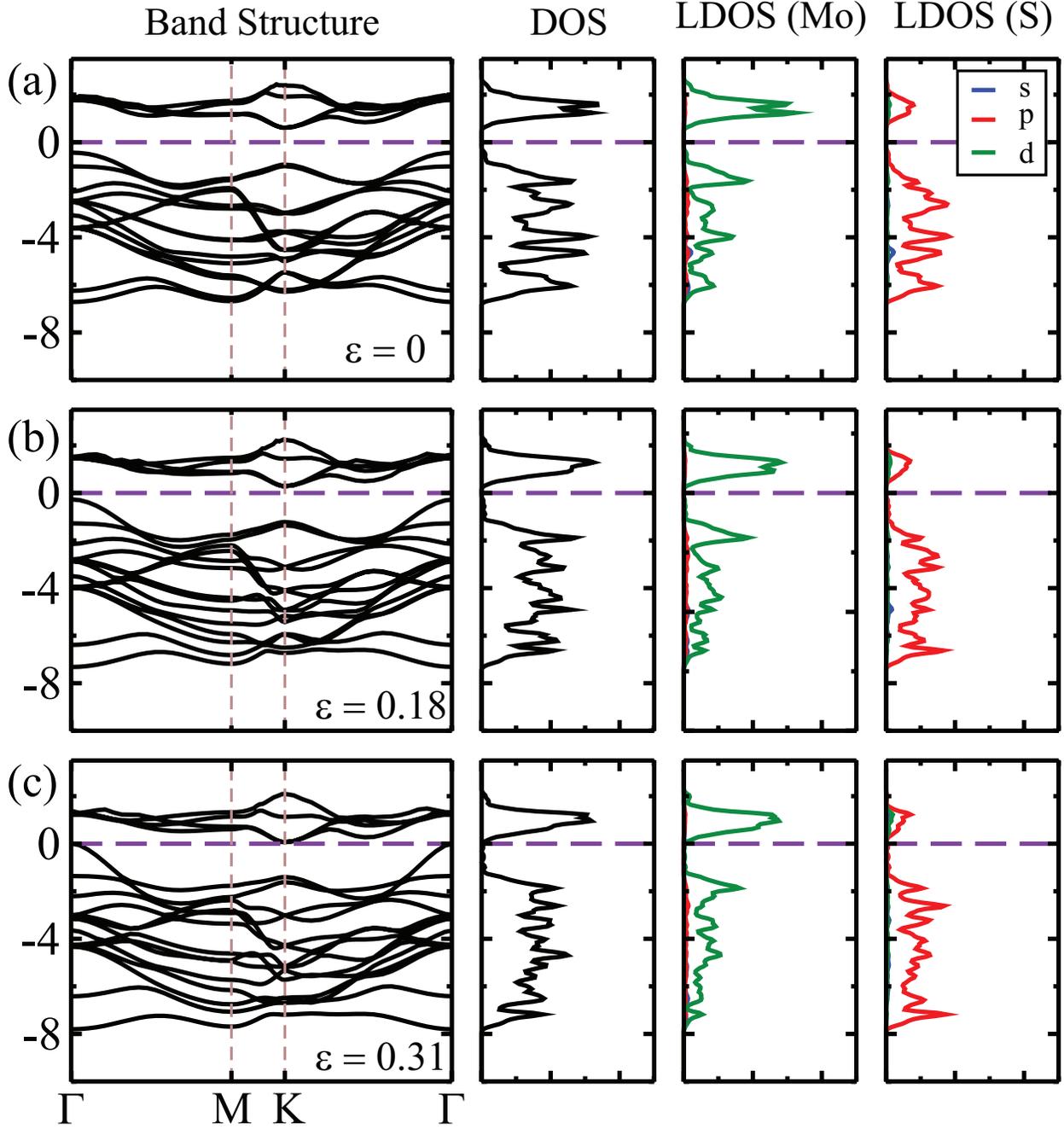}
\caption{ (Color online) Band structure, total DOS, PDOS of Mo and S for MoS$_2$ bilayer at (a) 0, (b) 0.18 and (c) 0.34 strain.}
\label{fig:2}
\end{figure}

\section{Results and discussion}

The two layers in a bilayer TMDs, can be arranged in AA and AB stacking. In the AA (AB) stacking the Mo atoms in one layer is on top of the Mo (S) atoms in another layer, Fig.~\ref{fig:1}. AB-stacking is preferred in both bulk and bilayer TMDs. In the case of a bilayer TMD the energy differences between two types of stackings are very small. Therefore, in the current work, the effect of applied normal compressive strain on the band structure of both AA- and AB- stacked TMDs, was studied. The strain was calculated as $\epsilon = (d_0-d)/d_0$, where $d_0$ and d are equilibrium and instantaneous interlayer distance, respectively. A constrained relaxation scheme, where the positions of metal atoms remained unchanged, was used for the optimization of the structures. This prevented the relaxation of instantaneous structures to the original positions. All the reported results here onwards include the vdW interactions via DFT-D2 method, unless mentioned otherwise.

The overall behavior of the band structure under the applied pressure remains similar for the both type of stackings. Therefore, here we will present mainly the results of AB-stacked TMDs. The PBE+vdW band structure calculations show that the unstrained bilayers of all these AB stacked TMDs are indirect band gap semiconductors. In comparison to monolayer, the band gap of bilayer decreases slightly and each band becomes doubly degenerated as shown in Fig.~\ref{fig:2} for MoS$_2$. This clearly demonstrates the absence of chemical interaction between the layers. As the interlayer separation decreases, the layers start to interact chemically, which leads to lifting of double degeneracy of the bands,  Fig.~\ref{fig:2}. This splitting increases with the increasing NCS. The valence band maxima (VBM) moves away/towards the fermi level at  K-/ $\Gamma$-point. The S-M transition occurs when VBM crosses the fermi level at the $\Gamma$ point after a critical applied NCS. Likewise, the conduction band minima (CBM) also moves towards the fermi level with increasing NCS. The band structures of the other MX$_2$ materials undergo similar changes (lifting of double degeneracy, and S-M transition) under applied NCS. The interlayer distance at which the S-M transition takes place is listed in Table~\ref{table:2} for the bilayers.
	
\begin{table}[b]
\centering
\begin{tabular}{c c c c c }
\hline
\hline
\multicolumn{1}{c}{Material }  & \multicolumn{2}{c}{ ~Bilayer separation at transition  (\AA)~} & 
\multicolumn{2}{c} {Transition Pressure (P$_{th}$) (GPa) } \\
\multicolumn{1}{c}{} & \multicolumn{1}{c}{AB} &
\multicolumn{1}{c}{ AA} &  \multicolumn{1}{c}{ AB}  & \multicolumn{1}{c}{ AA} \\
\hline
                    
 MoS$_2$    & 2.14 & 2.60 & 8.52, 14.47 (HSE)  & 8.37 \\
 MoSe$_2$    &  2.23 &2.64  &8.37 & 9.54  \\
 MoTe$_2$     &2.69  & 2.93 & 5.10 & 8.71 \\
 WS$_2$      & 1.80 & 2.47 & 16.28& 10.54  \\
  WSe$_2$     & 2.21 & 2.79 & 15.83 & 12.28 \\
\hline
\hline
\end{tabular}
\caption{Interlayer distance and pressure required for semiconductor to metal transition, P$_{th}$ of TMDs for both AA and AB stacking calculated using vdW.}
\label{table:2}
\end{table}

\begin{figure}[t]
\centering
\includegraphics[width=\columnwidth]{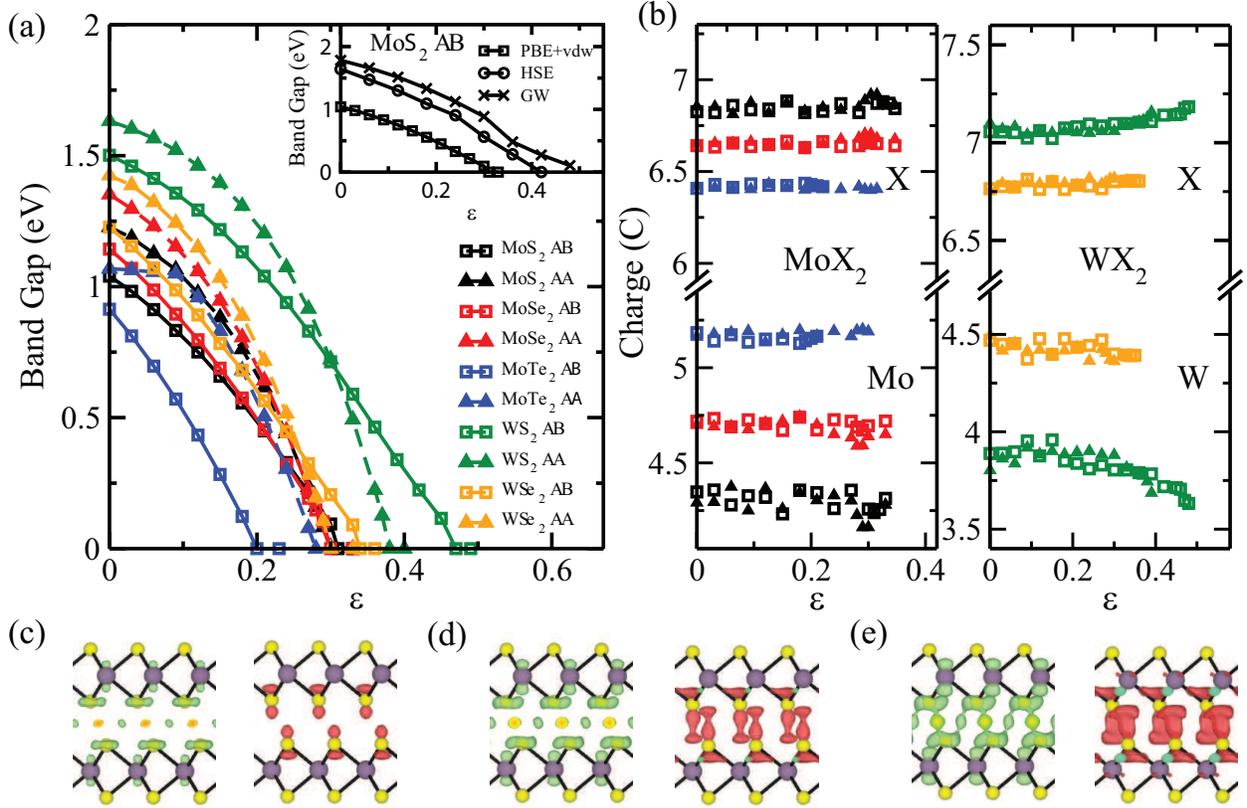}
\caption{ (Color online) (a) Change in band gap with applied strain for MX$_2$ bilayers for both AA and AB stacking using PBE+vdW method. The inset show the same using PBE+vdW, HSE and GW method for AB stacked MoS$_2$ bilayer. (b) Average Bader charge of M and X for all the bilayers as a function of strain. The left and right panels corresponds to M = Mo and W, respectively. Same color and symbols are used for the respective bilayers as defined in (a). Isosurfaces of charge accumulation (green)  and depletion (red) of MoS$_2$ bilayer under strain of  (c) 0.18, (d) 0.24 and (e) 0.31.}
\label{fig:3}
\end{figure}

In order to determine the constituents of the electronic bands, the total and projected density of states (PDOS) were calculated for the bilayers under various NCS and shown for MoS$_2$ in Fig.~\ref{fig:2}. The total DOS shows a gap near the fermi level, which reduces due to the shift of the CBM and the VBM towards the fermi level with increase in strain. For the unstrained MoS$_2$, the CBM and the VBM are constituted by Mo-$d$ and S-$p$ orbitals, which can be clearly seen from PDOS. With the increasing NCS, the contributions to CBM from Mo-$d$ and S-$p$ orbitals decrease, while from S-$d$ increase. The change in the band gap as a function of applied NCS follows similar pattern for all the materials studied here, (Fig.~\ref{fig:3}(a)). This change is reversible, i.e in the absence of the applied pressure the structure relaxes back to the original structure with the complete recovery of band gap, which is very important for sensor applications. Similar changes in band gap \cite {Wei2010} and resistance \cite {Dave2004} were reported for bulk MoS$_2$ under applied pressure, however, unlike bilayers no S-M transition was observed even at very high pressure (40 GPa). 

\begin{figure}[t]
\centering
\includegraphics[width=\columnwidth]{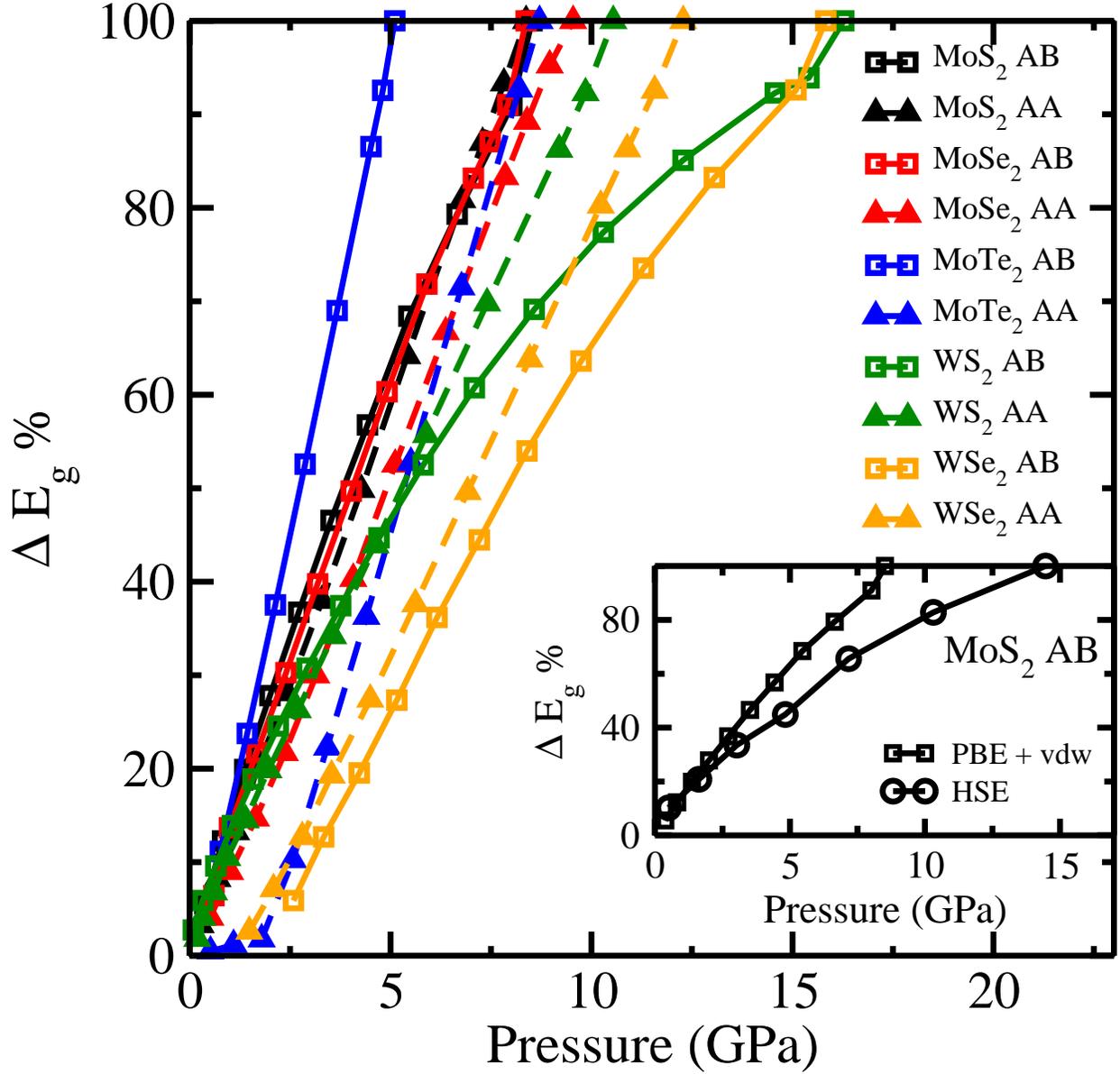}
\caption{ (Color online) Percentage change in band gap with applied pressure for MX$_2$ bilayers obtained using PBE-vdW. Inset:  The same plot using PBE-vdW and HSE method for AB stacked MoS$_2$ bilayer.}
\label{fig:4}
\end{figure}

In order to access the feasibility of S-M transition in experiments, the applied pressure (P) was calculated from the energy cost per unit area in reducing the interlayer distance by $\Delta d=(d_0 - d)$  as per the following equation,
\begin{equation}
P = \frac{E - E_0}{(d_0 - d)A} 
\label{eq:pressure}
\end{equation}

where A is the area of the unit cell, E and E$_0$ are the energies and d and d$_0$ are the interlayer distances of the strained and unstrained bilayer. The pressure required for S-M transition, P$_{th}$ for each material is listed in Table~\ref{table:2}. The calculated pressure range is easily achievable experimentally, hence makes tuning of band gap via NCS very attractive for various applications. 
Percentage reduction in band gap with increase in pressure is plotted  in Fig.~\ref{fig:4}. The plot is linear for most of the bilayers except for AB stacked WX$_2$. For MoTe$_2$, the change in band gap is very less initially but increases drastically after a pressure of 1.3 GPa. The slope and hence the response of change in band gap to applied pressure  for MoTe$_2$ (AB) is the largest among all the TMDs. Surprisingly, S-M transition occurs at relatively lower pressure for the AA stacking than the AB except for MoTe$_2$. For a given metal atom, the S-M transition pressure decreases as we go down the column of X atom in the periodic table. This is caused by increased delocalization of atomic orbitals, which leads to reduced interaction between M and X atoms resulting in S-M transition at lower pressure. Such behavior is consistent with the trend of band gap, which also decreases from S to Te, Table~\ref{table:1}. 

Similar evidence also comes from the Bader charge analysis \cite {Tang2009, Sanville2007, Henkelman2006}. The average charge on M and X as a function of applied strain is shown in Fig.~\ref{fig:3}(b). The charge transfer from M to X increases as a function of NCS, which is more prominent for WX$_2$. Furthermore, for a given M atom, the value of average charge on X atom decreases as we go down the periodic table i.e., from S to Te, replicating closely the trend of electron affinity of X atoms. This, essentially, results into weaker interaction between M-X atom, which leads to the S-M transitions pressure to be the lowest for the Te and the highest for the S.  Furthermore, in comparison to W, Mo has lower ionization potential, which measures the ability to donate the charge. Mo can donate charges relatively, easily and thereby facilitate the X-X interaction, which results in S-M transition in MoX$_2$ at lower pressure compared with the WX$_2$.

These conclusions were further supported by redistribution of charges,  which was calculated by taking the difference of the total charge of bilayer and  two isolated layers for MoS$_2$, as shown in Fig.~\ref{fig:3} c-e. The maximum redistribution of the charge occurs around the inner S atoms and the adjacent Mo-S bonds. As expected, for low strains (below $\epsilon = 0.21$) no charge rearrangement was observed. At the non zero strains, the charge started accumulating on the inner S atoms as well as on adjacent Mo-S bonds, Fig.~\ref{fig:3} c-e. However, the charge depletion was observed predominantly from the inner S atoms. With the increasing NCS, the amount of charge redistribution also increases, indicating the enhanced S-S interaction, which essentially causes the S-M transition. Due to smaller S-S distances the interaction is better for the AA-stacking compared with the AB. This is why the P$_{th}$ for AA-stacking is lowered compared with the AB-stacking.

\begin{figure}[t]
\centering
\includegraphics[width=\columnwidth]{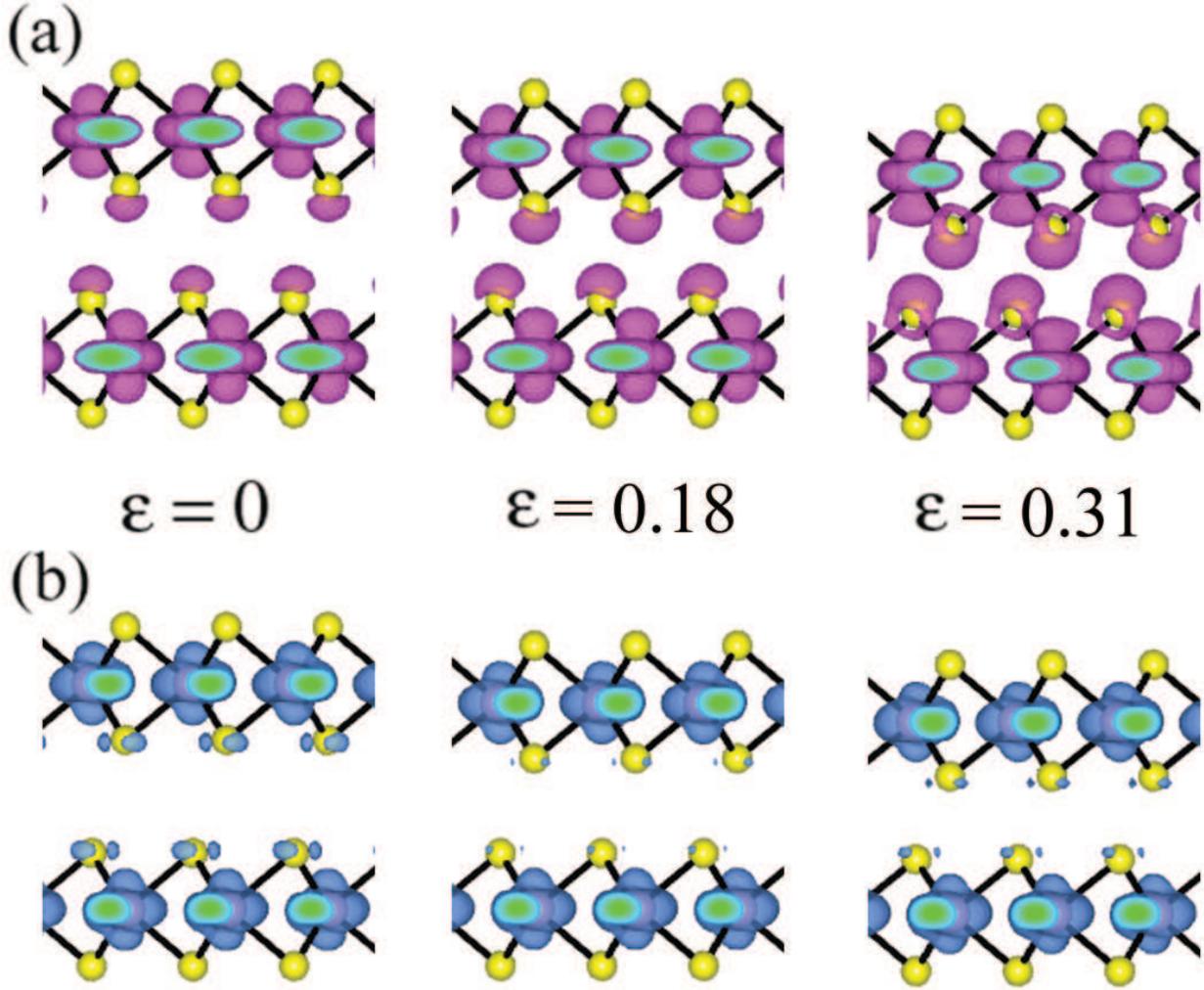}
\caption{ (Color online) Isosurfaces of band decomposed charge density of (a) valence band maxima and (b) conduction band minima of  AB staked MoS$_2$ bilayer at applied strain of 0, 0.18 and 0.31, respectively. }
\label{fig:5}
\end{figure}

In order to gain further insight, the band decomposed electron densities of the VBM and CBM, at different strains are plotted and shown in Fig.~\ref{fig:5}. Consistent with the above analysis, the VBM originates from the interaction of inner S atoms and adjacent Mo-S bonds,  Fig.~\ref{fig:5} a. With the increasing interaction caused by applied NCS, the contribution from inner S atoms further increases. The VBM is extended and hence the effective mass does not change much even after the application of NCS, implying maintenance of quality of conductivity. Furthermore the lowering of the symmetry of VBM is also evident. The CBM completely originates from the Mo-S bonds, Fig.~\ref{fig:5} b. Once again the nature is more extended and hence will keep the conductivities intact. The modification of gap without significant alteration of the dispersion of gap, is very importance for electronic applications.   

Due to the presence of artificial self-interaction\cite{Perdew1981} and the absence of the derivative discontinuity in the exchange-correlation potential, DFT in the local-density (LDA) and generalized gradient approximation (GGA), suffers from the underestimation of the band gap. The calculated PBE/GGA band gap of bulk, mono- and bi- layer MoS$_2$ is underestimated by an amount of 0.29 eV(23.7\%), 0.22 eV(11.5\%) and 0.01 eV(0.5\%) than their experimental values, respectively (Table~\ref{table:1}). The excellent agreement for the bi-layer is misleading as it corresponds to a wrong geometry (extremely large c value) obtained by PBE in the absence of vdW interactions.  Band gap of a bilayer with the geometry optimized by PBE+vdW is underestimated by 0.33 eV(20.6\%) with respect to the experimental value. For the comparison,  band gap using LDA functional was also calculated for MoS$_2$. The LDA predicts correct band gap for monolayer but gives values 0.47 eV(38.4\%) and 0.66 eV(41.1\%) lower than the experimental ones for the bulk and bilayer MoS$_2$ respectively. 

In order to correct the GGA/LDA band-gaps, we used hybrid Heyd-Scuseria-Ernzerhof (HSE) \cite{heyd2003, heyd2006} functional. In the HSE approach, the exchange potential is separated into a long-range and a short- range part.  The 1/4 of the PBE exchange is replaced by the Hartree-Fock (HF) exact exchange and the full PBE correlation energy is added. HSE is shown to correct the GGA band gaps \cite{heyd2003, Matsushita2011}, significantly by partially correcting the self interaction. However, the results are system dependent. The calculated band gaps for MoS$_2$ bilayer using HSE is 1.64 eV which is in very good agreement with the experiment (1.60 eV).   

We also corrected the PBE band gaps by using DFT with many-body perturbation theory in the GW approximation\cite{Hedin1965}. A partially self consistent GW0 method was used in which the G was iterated but the W was kept fixed to the initial DFT (PBE) W0. A default cutoff (280.0 eV) for the wave function was used for the GW calculation. Convergence study was performed for number of bands, k-points and frequency grid points to achieve a convergence within 10-20 meV for the band gaps. A $\Gamma$-centered Monkhorst-Pack k-point grid of $18 \times 18 \times 1$ mesh was used for calculating GW band structure. Like HSE, the GW0 also improves PBE band gap and overestimates it by 0.18 eV(11.3\%) (for bilayer) compared with experimental value. The GW0 band gap is slightly larger than the HSE. There are not many experimental results for the band gaps of the 2D-materials. Furthermore, the quality of as grown 2D-sheets can also play an important role in band-gap determination. In order to get a more reliable comparison more experiments are required. The band gap calculated using HSE  functional is the closest to the experimental value among all the methods used in this work for bilayer MoS$_2$.  Both HSE and GW calculations are computationally very expensive and hence we applied these methods only to MoS$_2$.

We checked the robustness of the S-M transition by calculating the band-structure of MoS$_2$ as a function of normal compressive strain using HSE functional and GW method. The plot of HSE and GW band gap as a function of applied strain is compared with the one for PBE+vdW  in Fig.~\ref{fig:3} a inset. A comparison of percentage reduction in band gap with increase in pressure for different materials and methods is shown in Fig.~\ref{fig:4}. Although, the calculated HSE and GW band gap of MoS$_2$ was slightly larger than the PBE, but the nature of the band structure as well as the S-M transition remained unchanged.  GW method does not give any force or energy of the systems and therefore pressure could not be calculated and compared with other methods for it.  Since the band gap obtained from HSE and GW calculation was slightly more than the ones with the GGA, an increase in transition interlayer distance (as well as P$_{th}$ for HSE) was observed for the S-M transition. The excellent agreement between the HSE and experimental band-gap, suggests that the transition temperature would be closer the HSE values. However, more experiments are required to claim a good numerical accuracy.  

\section {conclusion}

In conclusion, we report reversible band gap engineering of semiconducting bilayer of transition metal dichalcogenides by applied normal compressive strain. The band gap can be tuned in a large energy range without modifying the conductivity significantly. Furthermore, a universal reversible semiconductor to metal transition was observed for all the semiconducting TMDs. The reduction of gap as a function of applied pressure is caused by interlayer interaction, which eventually lifts the degeneracies of bands and moves them closer to the fermi level. For a given M in MX$_2$, the threshold pressure P$_{th}$, to achieve S-M transition decreases as X changes from S to Te. Furthermore, MoX$_2$ has lower P$_{th}$, compared with the WX$_2$. The PBE band gap of MoS$_{2}$ was corrected using hybrid HSE functional and GW method. The S-M transition can be reproduced even after applying the band gap corrections, however, the P$_{th}$ changes to a larger value. The tantalizing possibility of reversible tuning of more than 1.9 eV of energy gap by applying compressive strain as shown in the present work, would make the usage of TMDs in wide range of applications spanning from sensors to electronics.



\begin{thebibliography}{56}%
\makeatletter
\providecommand \@ifxundefined [1]{%
 \@ifx{#1\undefined}
}%
\providecommand \@ifnum [1]{%
 \ifnum #1\expandafter \@firstoftwo
 \else \expandafter \@secondoftwo
 \fi
}%
\providecommand \@ifx [1]{%
 \ifx #1\expandafter \@firstoftwo
 \else \expandafter \@secondoftwo
 \fi
}%
\providecommand \natexlab [1]{#1}%
\providecommand \enquote  [1]{``#1''}%
\providecommand \bibnamefont  [1]{#1}%
\providecommand \bibfnamefont [1]{#1}%
\providecommand \citenamefont [1]{#1}%
\providecommand \href@noop [0]{\@secondoftwo}%
\providecommand \href [0]{\begingroup \@sanitize@url \@href}%
\providecommand \@href[1]{\@@startlink{#1}\@@href}%
\providecommand \@@href[1]{\endgroup#1\@@endlink}%
\providecommand \@sanitize@url [0]{\catcode `\\12\catcode `\$12\catcode
  `\&12\catcode `\#12\catcode `\^12\catcode `\_12\catcode `\%12\relax}%
\providecommand \@@startlink[1]{}%
\providecommand \@@endlink[0]{}%
\providecommand \url  [0]{\begingroup\@sanitize@url \@url }%
\providecommand \@url [1]{\endgroup\@href {#1}{\urlprefix }}%
\providecommand \urlprefix  [0]{URL }%
\providecommand \Eprint [0]{\href }%
\@ifxundefined \urlstyle {%
  \providecommand \doi  [0]{\begingroup \@sanitize@url \@doi}%
  \providecommand \@doi [1]{\endgroup \@@startlink {\doibase
  #1}doi:\discretionary {}{}{}#1\@@endlink }%
}{%
  \providecommand \doi  [0]{doi:\discretionary{}{}{}\begingroup
  \urlstyle{rm}\Url }%
}%
\providecommand \doibase [0]{http://dx.doi.org/}%
\providecommand \Doi [0]{\begingroup \@sanitize@url \@Doi }%
\providecommand \@Doi  [1]{\endgroup\@@startlink{\doibase#1}\@@Doi}%
\providecommand \@@Doi [1]{#1\@@endlink}%
\providecommand \selectlanguage [0]{\@gobble}%
\providecommand \bibinfo  [0]{\@secondoftwo}%
\providecommand \bibfield  [0]{\@secondoftwo}%
\providecommand \translation [1]{[#1]}%
\providecommand \BibitemOpen [0]{}%
\providecommand \bibitemStop [0]{}%
\providecommand \bibitemNoStop [0]{.\EOS\space}%
\providecommand \EOS [0]{\spacefactor3000\relax}%
\providecommand \BibitemShut  [1]{\csname bibitem#1\endcsname}%
\bibitem [{\citenamefont {Ding}\ \emph {et~al.}(2011)\citenamefont {Ding},
  \citenamefont {Wang}, \citenamefont {Ni}, \citenamefont {Shi}, \citenamefont
  {Shi},\ and\ \citenamefont {Tang}}]{DingMX22011}%
  \BibitemOpen
  \bibfield  {author} {\bibinfo {author} {\bibfnamefont {Y.}~\bibnamefont
  {Ding}}, \bibinfo {author} {\bibfnamefont {Y.}~\bibnamefont {Wang}}, \bibinfo
  {author} {\bibfnamefont {J.}~\bibnamefont {Ni}}, \bibinfo {author}
  {\bibfnamefont {L.}~\bibnamefont {Shi}}, \bibinfo {author} {\bibfnamefont
  {S.}~\bibnamefont {Shi}}, \ and\ \bibinfo {author} {\bibfnamefont
  {W.}~\bibnamefont {Tang}},\ }\Doi {10.1016/j.physb.2011.03.044} {\bibfield
  {journal} {\bibinfo  {journal} {Physica B},\ }\textbf {\bibinfo {volume}
  {406}},\ \bibinfo {pages} {2254 } (\bibinfo {year} {2011})}\BibitemShut
  {NoStop}%
\bibitem [{\citenamefont {Ayari}\ \emph {et~al.}(2007)\citenamefont {Ayari},
  \citenamefont {Cobas}, \citenamefont {Ogundadegbe},\ and\ \citenamefont
  {Fuhrer}}]{ayari2007}%
  \BibitemOpen
  \bibfield  {author} {\bibinfo {author} {\bibfnamefont {A.}~\bibnamefont
  {Ayari}}, \bibinfo {author} {\bibfnamefont {E.}~\bibnamefont {Cobas}},
  \bibinfo {author} {\bibfnamefont {O.}~\bibnamefont {Ogundadegbe}}, \ and\
  \bibinfo {author} {\bibfnamefont {M.~S.}\ \bibnamefont {Fuhrer}},\ }\Doi
  {10.1063/1.2407388} {\bibfield  {journal} {\bibinfo  {journal} {J. Appl.
  Phys.},\ }\textbf {\bibinfo {volume} {101}},\ \bibinfo {eid} {014507}
  (\bibinfo {year} {2007})}\BibitemShut {NoStop}%
\bibitem [{\citenamefont {Gabovich}\ \emph {et~al.}(2001)\citenamefont
  {Gabovich}, \citenamefont {Voitenko}, \citenamefont {Annett},\ and\
  \citenamefont {Ausloos}}]{Gabovich2001}%
  \BibitemOpen
  \bibfield  {author} {\bibinfo {author} {\bibfnamefont {A.~M.}\ \bibnamefont
  {Gabovich}}, \bibinfo {author} {\bibfnamefont {A.~I.}\ \bibnamefont
  {Voitenko}}, \bibinfo {author} {\bibfnamefont {J.~F.}\ \bibnamefont
  {Annett}}, \ and\ \bibinfo {author} {\bibfnamefont {M.}~\bibnamefont
  {Ausloos}},\ }\href {http://stacks.iop.org/0953-2048/14/i=4/a=201} {\bibfield
   {journal} {\bibinfo  {journal} {Supercond. Sci. Tech.},\ }\textbf {\bibinfo
  {volume} {14}},\ \bibinfo {pages} {R1} (\bibinfo {year} {2001})}\BibitemShut
  {NoStop}%
\bibitem [{\citenamefont {Morosan}\ \emph {et~al.}(2006)\citenamefont
  {Morosan}, \citenamefont {Zandbergen}, \citenamefont {Dennis}, \citenamefont
  {Bos}, \citenamefont {Onose}, \citenamefont {Klimczuk}, \citenamefont
  {Ramirez}, \citenamefont {Ong},\ and\ \citenamefont {Cava}}]{Morosan2006}%
  \BibitemOpen
  \bibfield  {author} {\bibinfo {author} {\bibfnamefont {E.}~\bibnamefont
  {Morosan}}, \bibinfo {author} {\bibfnamefont {H.~W.}\ \bibnamefont
  {Zandbergen}}, \bibinfo {author} {\bibfnamefont {B.~S.}\ \bibnamefont
  {Dennis}}, \bibinfo {author} {\bibfnamefont {J.~W.~G.}\ \bibnamefont {Bos}},
  \bibinfo {author} {\bibfnamefont {Y.}~\bibnamefont {Onose}}, \bibinfo
  {author} {\bibfnamefont {T.}~\bibnamefont {Klimczuk}}, \bibinfo {author}
  {\bibfnamefont {A.~P.}\ \bibnamefont {Ramirez}}, \bibinfo {author}
  {\bibfnamefont {N.~P.}\ \bibnamefont {Ong}}, \ and\ \bibinfo {author}
  {\bibfnamefont {R.~J.}\ \bibnamefont {Cava}},\ }\href
  {http://dx.doi.org/10.1038/nphys360} {\bibfield  {journal} {\bibinfo
  {journal} {Nat. Phys.},\ }\textbf {\bibinfo {volume} {2}},\ \bibinfo {pages}
  {544} (\bibinfo {year} {2006})}\BibitemShut {NoStop}%
\bibitem [{\citenamefont {Rossnagel}\ \emph {et~al.}(2001)\citenamefont
  {Rossnagel}, \citenamefont {Seifarth}, \citenamefont {Kipp}, \citenamefont
  {Skibowski}, \citenamefont {Vo\ss{}}, \citenamefont {Kr\"uger}, \citenamefont
  {Mazur},\ and\ \citenamefont {Pollmann}}]{Rossnagel2001}%
  \BibitemOpen
  \bibfield  {author} {\bibinfo {author} {\bibfnamefont {K.}~\bibnamefont
  {Rossnagel}}, \bibinfo {author} {\bibfnamefont {O.}~\bibnamefont {Seifarth}},
  \bibinfo {author} {\bibfnamefont {L.}~\bibnamefont {Kipp}}, \bibinfo {author}
  {\bibfnamefont {M.}~\bibnamefont {Skibowski}}, \bibinfo {author}
  {\bibfnamefont {D.}~\bibnamefont {Vo\ss{}}}, \bibinfo {author} {\bibfnamefont
  {P.}~\bibnamefont {Kr\"uger}}, \bibinfo {author} {\bibfnamefont
  {A.}~\bibnamefont {Mazur}}, \ and\ \bibinfo {author} {\bibfnamefont
  {J.}~\bibnamefont {Pollmann}},\ }\Doi {10.1103/PhysRevB.64.235119} {\bibfield
   {journal} {\bibinfo  {journal} {Phys. Rev. B},\ }\textbf {\bibinfo {volume}
  {64}},\ \bibinfo {pages} {235119} (\bibinfo {year} {2001})}\BibitemShut
  {NoStop}%
\bibitem [{\citenamefont {Hu}\ \emph {et~al.}(2007)\citenamefont {Hu},
  \citenamefont {Li}, \citenamefont {Yan}, \citenamefont {Wen}, \citenamefont
  {Wu}, \citenamefont {Chen},\ and\ \citenamefont {Wang}}]{Hu2007}%
  \BibitemOpen
  \bibfield  {author} {\bibinfo {author} {\bibfnamefont {W.~Z.}\ \bibnamefont
  {Hu}}, \bibinfo {author} {\bibfnamefont {G.}~\bibnamefont {Li}}, \bibinfo
  {author} {\bibfnamefont {J.}~\bibnamefont {Yan}}, \bibinfo {author}
  {\bibfnamefont {H.~H.}\ \bibnamefont {Wen}}, \bibinfo {author} {\bibfnamefont
  {G.}~\bibnamefont {Wu}}, \bibinfo {author} {\bibfnamefont {X.~H.}\
  \bibnamefont {Chen}}, \ and\ \bibinfo {author} {\bibfnamefont {N.~L.}\
  \bibnamefont {Wang}},\ }\Doi {10.1103/PhysRevB.76.045103} {\bibfield
  {journal} {\bibinfo  {journal} {Phys. Rev. B},\ }\textbf {\bibinfo {volume}
  {76}},\ \bibinfo {pages} {045103} (\bibinfo {year} {2007})}\BibitemShut
  {NoStop}%
\bibitem [{\citenamefont {Kusmartseva}\ \emph {et~al.}(2008)\citenamefont
  {Kusmartseva}, \citenamefont {Akrap}, \citenamefont {Berger}, \citenamefont
  {Forro},\ and\ \citenamefont {Tutis}}]{Kusmartseva2008}%
  \BibitemOpen
  \bibfield  {author} {\bibinfo {author} {\bibfnamefont {A.~F.}\ \bibnamefont
  {Kusmartseva}}, \bibinfo {author} {\bibfnamefont {A.}~\bibnamefont {Akrap}},
  \bibinfo {author} {\bibfnamefont {H.}~\bibnamefont {Berger}}, \bibinfo
  {author} {\bibfnamefont {L.}~\bibnamefont {Forro}}, \ and\ \bibinfo {author}
  {\bibfnamefont {E.}~\bibnamefont {Tutis}},\ }\href
  {http://dx.doi.org/10.1038/nmat2318} {\bibfield  {journal} {\bibinfo
  {journal} {Nat. Mater.},\ }\textbf {\bibinfo {volume} {7}},\ \bibinfo {pages}
  {960} (\bibinfo {year} {2008})}\BibitemShut {NoStop}%
\bibitem [{\citenamefont {Kam}\ and\ \citenamefont
  {Parkinson}(1982)}]{KamBGap1982}%
  \BibitemOpen
  \bibfield  {author} {\bibinfo {author} {\bibfnamefont {K.~K.}\ \bibnamefont
  {Kam}}\ and\ \bibinfo {author} {\bibfnamefont {B.~A.}\ \bibnamefont
  {Parkinson}},\ }\Doi {10.1021/j100393a010} {\bibfield  {journal} {\bibinfo
  {journal} {J. Phys. Chem.},\ }\textbf {\bibinfo {volume} {86}},\ \bibinfo
  {pages} {463} (\bibinfo {year} {1982})}\BibitemShut {NoStop}%
\bibitem [{\citenamefont {Han}\ \emph {et~al.}(2011)\citenamefont {Han},
  \citenamefont {Kwon}, \citenamefont {Kim}, \citenamefont {Ryu}, \citenamefont
  {Yun}, \citenamefont {Kim}, \citenamefont {Hwang}, \citenamefont {Kang},
  \citenamefont {Baik}, \citenamefont {Shin},\ and\ \citenamefont
  {Hong}}]{Han2011}%
  \BibitemOpen
  \bibfield  {author} {\bibinfo {author} {\bibfnamefont {S.~W.}\ \bibnamefont
  {Han}}, \bibinfo {author} {\bibfnamefont {H.}~\bibnamefont {Kwon}}, \bibinfo
  {author} {\bibfnamefont {S.~K.}\ \bibnamefont {Kim}}, \bibinfo {author}
  {\bibfnamefont {S.}~\bibnamefont {Ryu}}, \bibinfo {author} {\bibfnamefont
  {W.~S.}\ \bibnamefont {Yun}}, \bibinfo {author} {\bibfnamefont {D.~H.}\
  \bibnamefont {Kim}}, \bibinfo {author} {\bibfnamefont {J.~H.}\ \bibnamefont
  {Hwang}}, \bibinfo {author} {\bibfnamefont {J.-S.}\ \bibnamefont {Kang}},
  \bibinfo {author} {\bibfnamefont {J.}~\bibnamefont {Baik}}, \bibinfo {author}
  {\bibfnamefont {H.~J.}\ \bibnamefont {Shin}}, \ and\ \bibinfo {author}
  {\bibfnamefont {S.~C.}\ \bibnamefont {Hong}},\ }\Doi
  {10.1103/PhysRevB.84.045409} {\bibfield  {journal} {\bibinfo  {journal}
  {Phys. Rev. B},\ }\textbf {\bibinfo {volume} {84}},\ \bibinfo {pages}
  {045409} (\bibinfo {year} {2011})}\BibitemShut {NoStop}%
\bibitem [{\citenamefont {Frindt}\ and\ \citenamefont
  {Yoffe}(1963)}]{Frindt1963}%
  \BibitemOpen
  \bibfield  {author} {\bibinfo {author} {\bibfnamefont {R.~F.}\ \bibnamefont
  {Frindt}}\ and\ \bibinfo {author} {\bibfnamefont {A.~D.}\ \bibnamefont
  {Yoffe}},\ }\href {http://www.jstor.org/stable/2414446} {\bibfield  {journal}
  {\bibinfo  {journal} {Proc. R. Soc. Lon. Ser.-A},\ }\textbf {\bibinfo
  {volume} {273}},\ \bibinfo {pages} {69} (\bibinfo {year} {1963})}\BibitemShut
  {NoStop}%
\bibitem [{\citenamefont {Novoselov}\ \emph {et~al.}(2005)\citenamefont
  {Novoselov}, \citenamefont {Jiang}, \citenamefont {Schedin}, \citenamefont
  {Booth}, \citenamefont {Khotkevich}, \citenamefont {Morozov},\ and\
  \citenamefont {Geim}}]{Novoselov2D2005}%
  \BibitemOpen
  \bibfield  {author} {\bibinfo {author} {\bibfnamefont {K.~S.}\ \bibnamefont
  {Novoselov}}, \bibinfo {author} {\bibfnamefont {D.}~\bibnamefont {Jiang}},
  \bibinfo {author} {\bibfnamefont {F.}~\bibnamefont {Schedin}}, \bibinfo
  {author} {\bibfnamefont {T.~J.}\ \bibnamefont {Booth}}, \bibinfo {author}
  {\bibfnamefont {V.~V.}\ \bibnamefont {Khotkevich}}, \bibinfo {author}
  {\bibfnamefont {S.~V.}\ \bibnamefont {Morozov}}, \ and\ \bibinfo {author}
  {\bibfnamefont {A.~K.}\ \bibnamefont {Geim}},\ }\Doi
  {10.1073/pnas.0502848102} {\bibfield  {journal} {\bibinfo  {journal} {Proc.
  Natl. Acad. Sci. USA},\ }\textbf {\bibinfo {volume} {102}},\ \bibinfo {pages}
  {10451} (\bibinfo {year} {2005})}\BibitemShut {NoStop}%
\bibitem [{\citenamefont {Frindt}(1966)}]{frindtMoS1966}%
  \BibitemOpen
  \bibfield  {author} {\bibinfo {author} {\bibfnamefont {R.~F.}\ \bibnamefont
  {Frindt}},\ }\Doi {10.1063/1.1708627} {\bibfield  {journal} {\bibinfo
  {journal} {J. Appl. Phys.},\ }\textbf {\bibinfo {volume} {37}},\ \bibinfo
  {pages} {1928} (\bibinfo {year} {1966})}\BibitemShut {NoStop}%
\bibitem [{\citenamefont {Joensen}\ \emph {et~al.}(1986)\citenamefont
  {Joensen}, \citenamefont {Frindt},\ and\ \citenamefont
  {Morrison}}]{Joensen1986}%
  \BibitemOpen
  \bibfield  {author} {\bibinfo {author} {\bibfnamefont {P.}~\bibnamefont
  {Joensen}}, \bibinfo {author} {\bibfnamefont {R.~F.}\ \bibnamefont {Frindt}},
  \ and\ \bibinfo {author} {\bibfnamefont {S.~R.}\ \bibnamefont {Morrison}},\
  }\Doi {10.1016/0025-5408(86)90011-5} {\bibfield  {journal} {\bibinfo
  {journal} {Mater. Res. Bull.},\ }\textbf {\bibinfo {volume} {21}},\ \bibinfo
  {pages} {457} (\bibinfo {year} {1986})}\BibitemShut {NoStop}%
\bibitem [{\citenamefont {Schumacher}\ \emph {et~al.}(1993)\citenamefont
  {Schumacher}, \citenamefont {Scandella}, \citenamefont {Kruse},\ and\
  \citenamefont {Prins}}]{Schumacher1993}%
  \BibitemOpen
  \bibfield  {author} {\bibinfo {author} {\bibfnamefont {A.}~\bibnamefont
  {Schumacher}}, \bibinfo {author} {\bibfnamefont {L.}~\bibnamefont
  {Scandella}}, \bibinfo {author} {\bibfnamefont {N.}~\bibnamefont {Kruse}}, \
  and\ \bibinfo {author} {\bibfnamefont {R.}~\bibnamefont {Prins}},\ }\Doi
  {10.1016/0039-6028(93)90875-K} {\bibfield  {journal} {\bibinfo  {journal}
  {Surf. Sci.},\ }\textbf {\bibinfo {volume} {289}},\ \bibinfo {pages} {L595 }
  (\bibinfo {year} {1993})}\BibitemShut {NoStop}%
\bibitem [{\citenamefont {Coleman}\ \emph {et~al.}(2011)\citenamefont
  {Coleman}, \citenamefont {Lotya}, \citenamefont {OÕNeill}, \citenamefont
  {Bergin}, \citenamefont {King}, \citenamefont {Khan}, \citenamefont {Young},
  \citenamefont {Gaucher}, \citenamefont {De}, \citenamefont {Smith},
  \citenamefont {Shvets}, \citenamefont {Arora}, \citenamefont {Stanton},
  \citenamefont {Kim}, \citenamefont {Lee}, \citenamefont {Kim}, \citenamefont
  {Duesberg}, \citenamefont {Hallam}, \citenamefont {Boland}, \citenamefont
  {Wang}, \citenamefont {Donegan}, \citenamefont {Grunlan}, \citenamefont
  {Moriarty}, \citenamefont {Shmeliov}, \citenamefont {Nicholls}, \citenamefont
  {Perkins}, \citenamefont {Grieveson}, \citenamefont {Theuwissen},
  \citenamefont {McComb}, \citenamefont {Nellist},\ and\ \citenamefont
  {Nicolosi}}]{ColemanLiquid2011}%
  \BibitemOpen
  \bibfield  {author} {\bibinfo {author} {\bibfnamefont {J.~N.}\ \bibnamefont
  {Coleman}}, \bibinfo {author} {\bibfnamefont {M.}~\bibnamefont {Lotya}},
  \bibinfo {author} {\bibfnamefont {A.}~\bibnamefont {OÕNeill}}, \bibinfo
  {author} {\bibfnamefont {S.~D.}\ \bibnamefont {Bergin}}, \bibinfo {author}
  {\bibfnamefont {P.~J.}\ \bibnamefont {King}}, \bibinfo {author}
  {\bibfnamefont {U.}~\bibnamefont {Khan}}, \bibinfo {author} {\bibfnamefont
  {K.}~\bibnamefont {Young}}, \bibinfo {author} {\bibfnamefont
  {A.}~\bibnamefont {Gaucher}}, \bibinfo {author} {\bibfnamefont
  {S.}~\bibnamefont {De}}, \bibinfo {author} {\bibfnamefont {R.~J.}\
  \bibnamefont {Smith}}, \bibinfo {author} {\bibfnamefont {I.~V.}\ \bibnamefont
  {Shvets}}, \bibinfo {author} {\bibfnamefont {S.~K.}\ \bibnamefont {Arora}},
  \bibinfo {author} {\bibfnamefont {G.}~\bibnamefont {Stanton}}, \bibinfo
  {author} {\bibfnamefont {H.-Y.}\ \bibnamefont {Kim}}, \bibinfo {author}
  {\bibfnamefont {K.}~\bibnamefont {Lee}}, \bibinfo {author} {\bibfnamefont
  {G.~T.}\ \bibnamefont {Kim}}, \bibinfo {author} {\bibfnamefont {G.~S.}\
  \bibnamefont {Duesberg}}, \bibinfo {author} {\bibfnamefont {T.}~\bibnamefont
  {Hallam}}, \bibinfo {author} {\bibfnamefont {J.~J.}\ \bibnamefont {Boland}},
  \bibinfo {author} {\bibfnamefont {J.~J.}\ \bibnamefont {Wang}}, \bibinfo
  {author} {\bibfnamefont {J.~F.}\ \bibnamefont {Donegan}}, \bibinfo {author}
  {\bibfnamefont {J.~C.}\ \bibnamefont {Grunlan}}, \bibinfo {author}
  {\bibfnamefont {G.}~\bibnamefont {Moriarty}}, \bibinfo {author}
  {\bibfnamefont {A.}~\bibnamefont {Shmeliov}}, \bibinfo {author}
  {\bibfnamefont {R.~J.}\ \bibnamefont {Nicholls}}, \bibinfo {author}
  {\bibfnamefont {J.~M.}\ \bibnamefont {Perkins}}, \bibinfo {author}
  {\bibfnamefont {E.~M.}\ \bibnamefont {Grieveson}}, \bibinfo {author}
  {\bibfnamefont {K.}~\bibnamefont {Theuwissen}}, \bibinfo {author}
  {\bibfnamefont {D.~W.}\ \bibnamefont {McComb}}, \bibinfo {author}
  {\bibfnamefont {P.~D.}\ \bibnamefont {Nellist}}, \ and\ \bibinfo {author}
  {\bibfnamefont {V.}~\bibnamefont {Nicolosi}},\ }\Doi
  {10.1126/science.1194975} {\bibfield  {journal} {\bibinfo  {journal}
  {Science},\ }\textbf {\bibinfo {volume} {331}},\ \bibinfo {pages} {568}
  (\bibinfo {year} {2011})}\BibitemShut {NoStop}%
\bibitem [{\citenamefont {Miremadi}\ and\ \citenamefont
  {Morrison}(1988)}]{miremadiWS1988}%
  \BibitemOpen
  \bibfield  {author} {\bibinfo {author} {\bibfnamefont {B.~K.}\ \bibnamefont
  {Miremadi}}\ and\ \bibinfo {author} {\bibfnamefont {S.~R.}\ \bibnamefont
  {Morrison}},\ }\Doi {10.1063/1.340441} {\bibfield  {journal} {\bibinfo
  {journal} {J. Appl. Phys.},\ }\textbf {\bibinfo {volume} {63}},\ \bibinfo
  {pages} {4970} (\bibinfo {year} {1988})}\BibitemShut {NoStop}%
\bibitem [{\citenamefont {Yang}\ and\ \citenamefont
  {Frindt}(1996)}]{YangWS21996}%
  \BibitemOpen
  \bibfield  {author} {\bibinfo {author} {\bibfnamefont {D.}~\bibnamefont
  {Yang}}\ and\ \bibinfo {author} {\bibfnamefont {R.}~\bibnamefont {Frindt}},\
  }\Doi {doi:10.1016/0022-3697(95)00406-8} {\bibfield  {journal} {\bibinfo
  {journal} {J. Phys. Chem. Solids},\ }\textbf {\bibinfo {volume} {57}},\
  \bibinfo {pages} {1113} (\bibinfo {year} {1996})}\BibitemShut {NoStop}%
\bibitem [{\citenamefont {Ramakrishna~Matte}\ \emph {et~al.}(2010)\citenamefont
  {Ramakrishna~Matte}, \citenamefont {Gomathi}, \citenamefont {Manna},
  \citenamefont {Late}, \citenamefont {Datta}, \citenamefont {Pati},\ and\
  \citenamefont {Rao}}]{Ramak2010}%
  \BibitemOpen
  \bibfield  {author} {\bibinfo {author} {\bibfnamefont {H.~S.~S.}\
  \bibnamefont {Ramakrishna~Matte}}, \bibinfo {author} {\bibfnamefont
  {A.}~\bibnamefont {Gomathi}}, \bibinfo {author} {\bibfnamefont {A.~K.}\
  \bibnamefont {Manna}}, \bibinfo {author} {\bibfnamefont {D.~J.}\ \bibnamefont
  {Late}}, \bibinfo {author} {\bibfnamefont {R.}~\bibnamefont {Datta}},
  \bibinfo {author} {\bibfnamefont {S.~K.}\ \bibnamefont {Pati}}, \ and\
  \bibinfo {author} {\bibfnamefont {C.~N.~R.}\ \bibnamefont {Rao}},\ }\Doi
  {10.1002/anie.201000009} {\bibfield  {journal} {\bibinfo  {journal} {Angew.
  Chem. Int. Edit.},\ }\textbf {\bibinfo {volume} {49}},\ \bibinfo {pages}
  {4059} (\bibinfo {year} {2010})}\BibitemShut {NoStop}%
\bibitem [{\citenamefont {Eda}\ \emph {et~al.}(2011)\citenamefont {Eda},
  \citenamefont {Yamaguchi}, \citenamefont {Voiry}, \citenamefont {Fujita},
  \citenamefont {Chen},\ and\ \citenamefont {Chhowalla}}]{Eda2011}%
  \BibitemOpen
  \bibfield  {author} {\bibinfo {author} {\bibfnamefont {G.}~\bibnamefont
  {Eda}}, \bibinfo {author} {\bibfnamefont {H.}~\bibnamefont {Yamaguchi}},
  \bibinfo {author} {\bibfnamefont {D.}~\bibnamefont {Voiry}}, \bibinfo
  {author} {\bibfnamefont {T.}~\bibnamefont {Fujita}}, \bibinfo {author}
  {\bibfnamefont {M.}~\bibnamefont {Chen}}, \ and\ \bibinfo {author}
  {\bibfnamefont {M.}~\bibnamefont {Chhowalla}},\ }\Doi {10.1021/nl201874w}
  {\bibfield  {journal} {\bibinfo  {journal} {Nano Lett.},\ }\textbf {\bibinfo
  {volume} {11}},\ \bibinfo {pages} {5111} (\bibinfo {year}
  {2011})}\BibitemShut {NoStop}%
\bibitem [{\citenamefont {Korn}\ \emph {et~al.}(2011)\citenamefont {Korn},
  \citenamefont {Heydrich}, \citenamefont {Hirmer}, \citenamefont
  {Schmutzler},\ and\ \citenamefont {Schuller}}]{korn2011}%
  \BibitemOpen
  \bibfield  {author} {\bibinfo {author} {\bibfnamefont {T.}~\bibnamefont
  {Korn}}, \bibinfo {author} {\bibfnamefont {S.}~\bibnamefont {Heydrich}},
  \bibinfo {author} {\bibfnamefont {M.}~\bibnamefont {Hirmer}}, \bibinfo
  {author} {\bibfnamefont {J.}~\bibnamefont {Schmutzler}}, \ and\ \bibinfo
  {author} {\bibfnamefont {C.}~\bibnamefont {Schuller}},\ }\Doi
  {10.1063/1.3636402} {\bibfield  {journal} {\bibinfo  {journal} {Appl. Phys.
  Lett.},\ }\textbf {\bibinfo {volume} {99}},\ \bibinfo {eid} {102109}
  (\bibinfo {year} {2011})}\BibitemShut {NoStop}%
\bibitem [{\citenamefont {Leb\`egue}\ and\ \citenamefont
  {Eriksson}(2009)}]{Lebegue2009}%
  \BibitemOpen
  \bibfield  {author} {\bibinfo {author} {\bibfnamefont {S.}~\bibnamefont
  {Leb\`egue}}\ and\ \bibinfo {author} {\bibfnamefont {O.}~\bibnamefont
  {Eriksson}},\ }\Doi {10.1103/PhysRevB.79.115409} {\bibfield  {journal}
  {\bibinfo  {journal} {Phys. Rev. B},\ }\textbf {\bibinfo {volume} {79}},\
  \bibinfo {pages} {115409} (\bibinfo {year} {2009})}\BibitemShut {NoStop}%
\bibitem [{\citenamefont {Splendiani}\ \emph {et~al.}(2010)\citenamefont
  {Splendiani}, \citenamefont {Sun}, \citenamefont {Zhang}, \citenamefont {Li},
  \citenamefont {Kim}, \citenamefont {Chim}, \citenamefont {Galli},\ and\
  \citenamefont {Wang}}]{Splendiani2010}%
  \BibitemOpen
  \bibfield  {author} {\bibinfo {author} {\bibfnamefont {A.}~\bibnamefont
  {Splendiani}}, \bibinfo {author} {\bibfnamefont {L.}~\bibnamefont {Sun}},
  \bibinfo {author} {\bibfnamefont {Y.}~\bibnamefont {Zhang}}, \bibinfo
  {author} {\bibfnamefont {T.}~\bibnamefont {Li}}, \bibinfo {author}
  {\bibfnamefont {J.}~\bibnamefont {Kim}}, \bibinfo {author} {\bibfnamefont
  {C.-Y.}\ \bibnamefont {Chim}}, \bibinfo {author} {\bibfnamefont
  {G.}~\bibnamefont {Galli}}, \ and\ \bibinfo {author} {\bibfnamefont
  {F.}~\bibnamefont {Wang}},\ }\Doi {10.1021/nl903868w} {\bibfield  {journal}
  {\bibinfo  {journal} {Nano Lett.},\ }\textbf {\bibinfo {volume} {10}},\
  \bibinfo {pages} {1271} (\bibinfo {year} {2010})}\BibitemShut {NoStop}%
\bibitem [{\citenamefont {Ellis}\ \emph {et~al.}(2011)\citenamefont {Ellis},
  \citenamefont {Lucero},\ and\ \citenamefont {Scuseria}}]{ellis2011}%
  \BibitemOpen
  \bibfield  {author} {\bibinfo {author} {\bibfnamefont {J.~K.}\ \bibnamefont
  {Ellis}}, \bibinfo {author} {\bibfnamefont {M.~J.}\ \bibnamefont {Lucero}}, \
  and\ \bibinfo {author} {\bibfnamefont {G.~E.}\ \bibnamefont {Scuseria}},\
  }\Doi {10.1063/1.3672219} {\bibfield  {journal} {\bibinfo  {journal} {Appl.
  Phys. Lett.},\ }\textbf {\bibinfo {volume} {99}},\ \bibinfo {eid} {261908}
  (\bibinfo {year} {2011})}\BibitemShut {NoStop}%
\bibitem [{\citenamefont {Inosov}\ \emph {et~al.}(2008)\citenamefont {Inosov},
  \citenamefont {Zabolotnyy}, \citenamefont {Evtushinsky}, \citenamefont
  {Kordyuk}, \citenamefont {B\"{u}chner}, \citenamefont {Follath},
  \citenamefont {Berger},\ and\ \citenamefont {Borisenko}}]{Inosov2008}%
  \BibitemOpen
  \bibfield  {author} {\bibinfo {author} {\bibfnamefont {D.~S.}\ \bibnamefont
  {Inosov}}, \bibinfo {author} {\bibfnamefont {V.~B.}\ \bibnamefont
  {Zabolotnyy}}, \bibinfo {author} {\bibfnamefont {D.~V.}\ \bibnamefont
  {Evtushinsky}}, \bibinfo {author} {\bibfnamefont {A.~A.}\ \bibnamefont
  {Kordyuk}}, \bibinfo {author} {\bibfnamefont {B.}~\bibnamefont
  {B\"{u}chner}}, \bibinfo {author} {\bibfnamefont {R.}~\bibnamefont
  {Follath}}, \bibinfo {author} {\bibfnamefont {H.}~\bibnamefont {Berger}}, \
  and\ \bibinfo {author} {\bibfnamefont {S.~V.}\ \bibnamefont {Borisenko}},\
  }\href {http://stacks.iop.org/1367-2630/10/i=12/a=125027} {\bibfield
  {journal} {\bibinfo  {journal} {New J. Phys.},\ }\textbf {\bibinfo {volume}
  {10}},\ \bibinfo {pages} {125027} (\bibinfo {year} {2008})}\BibitemShut
  {NoStop}%
\bibitem [{\citenamefont {Kidd}\ \emph {et~al.}(2011)\citenamefont {Kidd},
  \citenamefont {Klein}, \citenamefont {Rash},\ and\ \citenamefont
  {Strauss}}]{Kidd2011}%
  \BibitemOpen
  \bibfield  {author} {\bibinfo {author} {\bibfnamefont {T.}~\bibnamefont
  {Kidd}}, \bibinfo {author} {\bibfnamefont {D.}~\bibnamefont {Klein}},
  \bibinfo {author} {\bibfnamefont {T.}~\bibnamefont {Rash}}, \ and\ \bibinfo
  {author} {\bibfnamefont {L.}~\bibnamefont {Strauss}},\ }\Doi
  {10.1016/j.apsusc.2010.11.156} {\bibfield  {journal} {\bibinfo  {journal}
  {Appl. Surf. Sci.},\ }\textbf {\bibinfo {volume} {257}},\ \bibinfo {pages}
  {3812 } (\bibinfo {year} {2011})}\BibitemShut {NoStop}%
\bibitem [{\citenamefont {Colev}\ \emph {et~al.}(2009)\citenamefont {Colev},
  \citenamefont {Gherman}, \citenamefont {Mirovitskii}, \citenamefont
  {Kulyuk},\ and\ \citenamefont {Fortin}}]{Colev2009}%
  \BibitemOpen
  \bibfield  {author} {\bibinfo {author} {\bibfnamefont {A.}~\bibnamefont
  {Colev}}, \bibinfo {author} {\bibfnamefont {C.}~\bibnamefont {Gherman}},
  \bibinfo {author} {\bibfnamefont {V.}~\bibnamefont {Mirovitskii}}, \bibinfo
  {author} {\bibfnamefont {L.}~\bibnamefont {Kulyuk}}, \ and\ \bibinfo {author}
  {\bibfnamefont {E.}~\bibnamefont {Fortin}},\ }\Doi
  {10.1016/j.jlumin.2009.05.022} {\bibfield  {journal} {\bibinfo  {journal} {J.
  Lumin.},\ }\textbf {\bibinfo {volume} {129}},\ \bibinfo {pages} {1945 }
  (\bibinfo {year} {2009})}\BibitemShut {NoStop}%
\bibitem [{\citenamefont {Radisavljevic}\ \emph {et~al.}(2010)\citenamefont
  {Radisavljevic}, \citenamefont {Radenovic}, \citenamefont {Brivio},
  \citenamefont {Giacometti},\ and\ \citenamefont
  {Kis}}]{RadisavljevicNatnano}%
  \BibitemOpen
  \bibfield  {author} {\bibinfo {author} {\bibfnamefont {B.}~\bibnamefont
  {Radisavljevic}}, \bibinfo {author} {\bibfnamefont {A.}~\bibnamefont
  {Radenovic}}, \bibinfo {author} {\bibfnamefont {J.}~\bibnamefont {Brivio}},
  \bibinfo {author} {\bibfnamefont {V.}~\bibnamefont {Giacometti}}, \ and\
  \bibinfo {author} {\bibfnamefont {A.}~\bibnamefont {Kis}},\ }\Doi
  {10.1038/nnano.2010.279} {\bibfield  {journal} {\bibinfo  {journal} {Nat.
  Nanotechnol.},\ }\textbf {\bibinfo {volume} {6}},\ \bibinfo {pages} {147}
  (\bibinfo {year} {2010})}\BibitemShut {NoStop}%
\bibitem [{\citenamefont {Liu}\ \emph {et~al.}(2011)\citenamefont {Liu},
  \citenamefont {Kumar}, \citenamefont {Ouyang},\ and\ \citenamefont
  {Guo}}]{Liu2011}%
  \BibitemOpen
  \bibfield  {author} {\bibinfo {author} {\bibfnamefont {L.}~\bibnamefont
  {Liu}}, \bibinfo {author} {\bibfnamefont {S.}~\bibnamefont {Kumar}}, \bibinfo
  {author} {\bibfnamefont {Y.}~\bibnamefont {Ouyang}}, \ and\ \bibinfo {author}
  {\bibfnamefont {J.}~\bibnamefont {Guo}},\ }\Doi {10.1109/TED.2011.2159221}
  {\bibfield  {journal} {\bibinfo  {journal} {IEEE T. Electron Dev.},\ }\textbf
  {\bibinfo {volume} {58}},\ \bibinfo {pages} {3042} (\bibinfo {year}
  {2011})}\BibitemShut {NoStop}%
\bibitem [{\citenamefont {Yoon}\ \emph {et~al.}(2011)\citenamefont {Yoon},
  \citenamefont {Ganapathi},\ and\ \citenamefont {Salahuddin}}]{Yoon2011}%
  \BibitemOpen
  \bibfield  {author} {\bibinfo {author} {\bibfnamefont {Y.}~\bibnamefont
  {Yoon}}, \bibinfo {author} {\bibfnamefont {K.}~\bibnamefont {Ganapathi}}, \
  and\ \bibinfo {author} {\bibfnamefont {S.}~\bibnamefont {Salahuddin}},\ }\Doi
  {10.1021/nl2018178} {\bibfield  {journal} {\bibinfo  {journal} {Nano Lett.},\
  }\textbf {\bibinfo {volume} {11}},\ \bibinfo {pages} {3768} (\bibinfo {year}
  {2011})}\BibitemShut {NoStop}%
\bibitem [{\citenamefont {Radisavljevic}\ \emph {et~al.}(2011)\citenamefont
  {Radisavljevic}, \citenamefont {Whitwick},\ and\ \citenamefont
  {Kis}}]{Radisavljevic2011}%
  \BibitemOpen
  \bibfield  {author} {\bibinfo {author} {\bibfnamefont {B.}~\bibnamefont
  {Radisavljevic}}, \bibinfo {author} {\bibfnamefont {M.~B.}\ \bibnamefont
  {Whitwick}}, \ and\ \bibinfo {author} {\bibfnamefont {A.}~\bibnamefont
  {Kis}},\ }\Doi {10.1021/nn203715c} {\bibfield  {journal} {\bibinfo  {journal}
  {ACS Nano},\ }\textbf {\bibinfo {volume} {5}},\ \bibinfo {pages} {9934}
  (\bibinfo {year} {2011})}\BibitemShut {NoStop}%
\bibitem [{\citenamefont {Ni}\ \emph {et~al.}(2008)\citenamefont {Ni},
  \citenamefont {Yu}, \citenamefont {Lu}, \citenamefont {Wang}, \citenamefont
  {Feng},\ and\ \citenamefont {Shen}}]{Zhen2008}%
  \BibitemOpen
  \bibfield  {author} {\bibinfo {author} {\bibfnamefont {Z.~H.}\ \bibnamefont
  {Ni}}, \bibinfo {author} {\bibfnamefont {T.}~\bibnamefont {Yu}}, \bibinfo
  {author} {\bibfnamefont {Y.~H.}\ \bibnamefont {Lu}}, \bibinfo {author}
  {\bibfnamefont {Y.~Y.}\ \bibnamefont {Wang}}, \bibinfo {author}
  {\bibfnamefont {Y.~P.}\ \bibnamefont {Feng}}, \ and\ \bibinfo {author}
  {\bibfnamefont {Z.~X.}\ \bibnamefont {Shen}},\ }\Doi {10.1021/nn800459e}
  {\bibfield  {journal} {\bibinfo  {journal} {ACS Nano},\ }\textbf {\bibinfo
  {volume} {2}},\ \bibinfo {pages} {2301} (\bibinfo {year} {2008})}\BibitemShut
  {NoStop}%
\bibitem [{\citenamefont {McCann}(2006)}]{McCann2006}%
  \BibitemOpen
  \bibfield  {author} {\bibinfo {author} {\bibfnamefont {E.}~\bibnamefont
  {McCann}},\ }\Doi {10.1103/PhysRevB.74.161403} {\bibfield  {journal}
  {\bibinfo  {journal} {Phys. Rev. B},\ }\textbf {\bibinfo {volume} {74}},\
  \bibinfo {pages} {161403} (\bibinfo {year} {2006})}\BibitemShut {NoStop}%
\bibitem [{\citenamefont {Castro}\ \emph {et~al.}(2007)\citenamefont {Castro},
  \citenamefont {Novoselov}, \citenamefont {Morozov}, \citenamefont {Peres},
  \citenamefont {Lopes~dos Santos}, \citenamefont {Nilsson}, \citenamefont
  {Guinea}, \citenamefont {Geim},\ and\ \citenamefont {Neto}}]{Castro2007}%
  \BibitemOpen
  \bibfield  {author} {\bibinfo {author} {\bibfnamefont {E.~V.}\ \bibnamefont
  {Castro}}, \bibinfo {author} {\bibfnamefont {K.~S.}\ \bibnamefont
  {Novoselov}}, \bibinfo {author} {\bibfnamefont {S.~V.}\ \bibnamefont
  {Morozov}}, \bibinfo {author} {\bibfnamefont {N.~M.~R.}\ \bibnamefont
  {Peres}}, \bibinfo {author} {\bibfnamefont {J.~M.~B.}\ \bibnamefont
  {Lopes~dos Santos}}, \bibinfo {author} {\bibfnamefont {J.}~\bibnamefont
  {Nilsson}}, \bibinfo {author} {\bibfnamefont {F.}~\bibnamefont {Guinea}},
  \bibinfo {author} {\bibfnamefont {A.~K.}\ \bibnamefont {Geim}}, \ and\
  \bibinfo {author} {\bibfnamefont {A.~H.~C.}\ \bibnamefont {Neto}},\ }\Doi
  {10.1103/PhysRevLett.99.216802} {\bibfield  {journal} {\bibinfo  {journal}
  {Phys. Rev. Lett.},\ }\textbf {\bibinfo {volume} {99}},\ \bibinfo {pages}
  {216802} (\bibinfo {year} {2007})}\BibitemShut {NoStop}%
\bibitem [{\citenamefont {Ohta}\ \emph {et~al.}(2006)\citenamefont {Ohta},
  \citenamefont {Bostwick}, \citenamefont {Seyller}, \citenamefont {Horn},\
  and\ \citenamefont {Rotenberg}}]{Ohta2006}%
  \BibitemOpen
  \bibfield  {author} {\bibinfo {author} {\bibfnamefont {T.}~\bibnamefont
  {Ohta}}, \bibinfo {author} {\bibfnamefont {A.}~\bibnamefont {Bostwick}},
  \bibinfo {author} {\bibfnamefont {T.}~\bibnamefont {Seyller}}, \bibinfo
  {author} {\bibfnamefont {K.}~\bibnamefont {Horn}}, \ and\ \bibinfo {author}
  {\bibfnamefont {E.}~\bibnamefont {Rotenberg}},\ }\Doi
  {10.1126/science.1130681} {\bibfield  {journal} {\bibinfo  {journal}
  {Science},\ }\textbf {\bibinfo {volume} {313}},\ \bibinfo {pages} {951}
  (\bibinfo {year} {2006})}\BibitemShut {NoStop}%
\bibitem [{\citenamefont {Zhang}\ \emph {et~al.}(2009)\citenamefont {Zhang},
  \citenamefont {Tang}, \citenamefont {Girit}, \citenamefont {Hao},
  \citenamefont {Martin}, \citenamefont {Zettl}, \citenamefont {Crommie},
  \citenamefont {Shen},\ and\ \citenamefont {Wang}}]{Zhang2009}%
  \BibitemOpen
  \bibfield  {author} {\bibinfo {author} {\bibfnamefont {Y.}~\bibnamefont
  {Zhang}}, \bibinfo {author} {\bibfnamefont {T.-T.}\ \bibnamefont {Tang}},
  \bibinfo {author} {\bibfnamefont {C.}~\bibnamefont {Girit}}, \bibinfo
  {author} {\bibfnamefont {Z.}~\bibnamefont {Hao}}, \bibinfo {author}
  {\bibfnamefont {M.~C.}\ \bibnamefont {Martin}}, \bibinfo {author}
  {\bibfnamefont {A.}~\bibnamefont {Zettl}}, \bibinfo {author} {\bibfnamefont
  {M.~F.}\ \bibnamefont {Crommie}}, \bibinfo {author} {\bibfnamefont {Y.~R.}\
  \bibnamefont {Shen}}, \ and\ \bibinfo {author} {\bibfnamefont
  {F.}~\bibnamefont {Wang}},\ }\Doi {10.1038/nature08105} {\bibfield  {journal}
  {\bibinfo  {journal} {Nature},\ }\textbf {\bibinfo {volume} {459}},\ \bibinfo
  {pages} {820} (\bibinfo {year} {2009})}\BibitemShut {NoStop}%
\bibitem [{\citenamefont {Xia}\ \emph {et~al.}(2010)\citenamefont {Xia},
  \citenamefont {Farmer}, \citenamefont {Lin},\ and\ \citenamefont
  {Avouris}}]{Xia2010}%
  \BibitemOpen
  \bibfield  {author} {\bibinfo {author} {\bibfnamefont {F.}~\bibnamefont
  {Xia}}, \bibinfo {author} {\bibfnamefont {D.~B.}\ \bibnamefont {Farmer}},
  \bibinfo {author} {\bibfnamefont {Y.-m.}\ \bibnamefont {Lin}}, \ and\
  \bibinfo {author} {\bibfnamefont {P.}~\bibnamefont {Avouris}},\ }\Doi
  {10.1021/nl9039636} {\bibfield  {journal} {\bibinfo  {journal} {Nano Lett.},\
  }\textbf {\bibinfo {volume} {10}},\ \bibinfo {pages} {715} (\bibinfo {year}
  {2010})}\BibitemShut {NoStop}%
\bibitem [{\citenamefont {Ramasubramaniam}\ \emph {et~al.}(2011)\citenamefont
  {Ramasubramaniam}, \citenamefont {Naveh},\ and\ \citenamefont
  {Towe}}]{Ramasubramaniam2011}%
  \BibitemOpen
  \bibfield  {author} {\bibinfo {author} {\bibfnamefont {A.}~\bibnamefont
  {Ramasubramaniam}}, \bibinfo {author} {\bibfnamefont {D.}~\bibnamefont
  {Naveh}}, \ and\ \bibinfo {author} {\bibfnamefont {E.}~\bibnamefont {Towe}},\
  }\Doi {10.1103/PhysRevB.84.205325} {\bibfield  {journal} {\bibinfo  {journal}
  {Phys. Rev. B},\ }\textbf {\bibinfo {volume} {84}},\ \bibinfo {pages}
  {205325} (\bibinfo {year} {2011})}\BibitemShut {NoStop}%
\bibitem [{\citenamefont {Scalise}\ \emph {et~al.}(2012)\citenamefont
  {Scalise}, \citenamefont {Houssa}, \citenamefont {Pourtois}, \citenamefont
  {Afanas'ev},\ and\ \citenamefont {Stesmans}}]{Scalise2012}%
  \BibitemOpen
  \bibfield  {author} {\bibinfo {author} {\bibfnamefont {E.}~\bibnamefont
  {Scalise}}, \bibinfo {author} {\bibfnamefont {M.}~\bibnamefont {Houssa}},
  \bibinfo {author} {\bibfnamefont {G.}~\bibnamefont {Pourtois}}, \bibinfo
  {author} {\bibfnamefont {V.}~\bibnamefont {Afanas'ev}}, \ and\ \bibinfo
  {author} {\bibfnamefont {A.}~\bibnamefont {Stesmans}},\ }\href
  {http://dx.doi.org/10.1007/s12274-011-0183-0} {\bibfield  {journal} {\bibinfo
   {journal} {Nano Res.},\ }\textbf {\bibinfo {volume} {5}},\ \bibinfo {pages}
  {43} (\bibinfo {year} {2012})}\BibitemShut {NoStop}%
\bibitem [{\citenamefont {Dave}\ \emph {et~al.}(2004)\citenamefont {Dave},
  \citenamefont {Vaidya}, \citenamefont {Patel},\ and\ \citenamefont
  {Jani}}]{Dave2004}%
  \BibitemOpen
  \bibfield  {author} {\bibinfo {author} {\bibfnamefont {M.}~\bibnamefont
  {Dave}}, \bibinfo {author} {\bibfnamefont {R.}~\bibnamefont {Vaidya}},
  \bibinfo {author} {\bibfnamefont {S.~G.}\ \bibnamefont {Patel}}, \ and\
  \bibinfo {author} {\bibfnamefont {A.~R.}\ \bibnamefont {Jani}},\ }\href@noop
  {} {\bibfield  {journal} {\bibinfo  {journal} {Bull. Mater. Sci.},\ }\textbf
  {\bibinfo {volume} {27}},\ \bibinfo {pages} {213} (\bibinfo {year}
  {2004})}\BibitemShut {NoStop}%
\bibitem [{\citenamefont {B\"oker}\ \emph {et~al.}(2001)\citenamefont
  {B\"oker}, \citenamefont {Severin}, \citenamefont {M\"uller}, \citenamefont
  {Janowitz}, \citenamefont {Manzke}, \citenamefont {Vo\ss{}}, \citenamefont
  {Kr\"uger}, \citenamefont {Mazur},\ and\ \citenamefont
  {Pollmann}}]{Boker2001}%
  \BibitemOpen
  \bibfield  {author} {\bibinfo {author} {\bibfnamefont {T.}~\bibnamefont
  {B\"oker}}, \bibinfo {author} {\bibfnamefont {R.}~\bibnamefont {Severin}},
  \bibinfo {author} {\bibfnamefont {A.}~\bibnamefont {M\"uller}}, \bibinfo
  {author} {\bibfnamefont {C.}~\bibnamefont {Janowitz}}, \bibinfo {author}
  {\bibfnamefont {R.}~\bibnamefont {Manzke}}, \bibinfo {author} {\bibfnamefont
  {D.}~\bibnamefont {Vo\ss{}}}, \bibinfo {author} {\bibfnamefont
  {P.}~\bibnamefont {Kr\"uger}}, \bibinfo {author} {\bibfnamefont
  {A.}~\bibnamefont {Mazur}}, \ and\ \bibinfo {author} {\bibfnamefont
  {J.}~\bibnamefont {Pollmann}},\ }\Doi {10.1103/PhysRevB.64.235305} {\bibfield
   {journal} {\bibinfo  {journal} {Phys. Rev. B},\ }\textbf {\bibinfo {volume}
  {64}},\ \bibinfo {pages} {235305} (\bibinfo {year} {2001})}\BibitemShut
  {NoStop}%
\bibitem [{\citenamefont {Mak}\ \emph {et~al.}(2010)\citenamefont {Mak},
  \citenamefont {Lee}, \citenamefont {Hone}, \citenamefont {Shan},\ and\
  \citenamefont {Heinz}}]{Mak2010}%
  \BibitemOpen
  \bibfield  {author} {\bibinfo {author} {\bibfnamefont {K.~F.}\ \bibnamefont
  {Mak}}, \bibinfo {author} {\bibfnamefont {C.}~\bibnamefont {Lee}}, \bibinfo
  {author} {\bibfnamefont {J.}~\bibnamefont {Hone}}, \bibinfo {author}
  {\bibfnamefont {J.}~\bibnamefont {Shan}}, \ and\ \bibinfo {author}
  {\bibfnamefont {T.~F.}\ \bibnamefont {Heinz}},\ }\Doi
  {10.1103/PhysRevLett.105.136805} {\bibfield  {journal} {\bibinfo  {journal}
  {Phys. Rev. Lett.},\ }\textbf {\bibinfo {volume} {105}},\ \bibinfo {pages}
  {136805} (\bibinfo {year} {2010})}\BibitemShut {NoStop}%
\bibitem [{\citenamefont {Schutte}\ \emph {et~al.}(1987)\citenamefont
  {Schutte}, \citenamefont {Boer},\ and\ \citenamefont
  {Jellinek}}]{Schutte1987}%
  \BibitemOpen
  \bibfield  {author} {\bibinfo {author} {\bibfnamefont {W.}~\bibnamefont
  {Schutte}}, \bibinfo {author} {\bibfnamefont {J.~D.}\ \bibnamefont {Boer}}, \
  and\ \bibinfo {author} {\bibfnamefont {F.}~\bibnamefont {Jellinek}},\ }\Doi
  {10.1016/0022-4596(87)90057-0} {\bibfield  {journal} {\bibinfo  {journal} {J.
  Solid State Chem.},\ }\textbf {\bibinfo {volume} {70}},\ \bibinfo {pages}
  {207 } (\bibinfo {year} {1987})}\BibitemShut {NoStop}%
\bibitem [{\citenamefont {Bl\"ochl}(1994)}]{Blochl94}%
  \BibitemOpen
  \bibfield  {author} {\bibinfo {author} {\bibfnamefont {P.~E.}\ \bibnamefont
  {Bl\"ochl}},\ }\Doi {10.1103/PhysRevB.50.17953} {\bibfield  {journal}
  {\bibinfo  {journal} {Phys. Rev. B},\ }\textbf {\bibinfo {volume} {50}},\
  \bibinfo {pages} {17953} (\bibinfo {year} {1994})}\BibitemShut {NoStop}%
\bibitem [{\citenamefont {Kresse}\ and\ \citenamefont
  {Joubert}(1999)}]{Kresse99}%
  \BibitemOpen
  \bibfield  {author} {\bibinfo {author} {\bibfnamefont {G.}~\bibnamefont
  {Kresse}}\ and\ \bibinfo {author} {\bibfnamefont {D.}~\bibnamefont
  {Joubert}},\ }\Doi {10.1103/PhysRevB.59.1758} {\bibfield  {journal} {\bibinfo
   {journal} {Phys. Rev. B},\ }\textbf {\bibinfo {volume} {59}},\ \bibinfo
  {pages} {1758} (\bibinfo {year} {1999})}\BibitemShut {NoStop}%
\bibitem [{\citenamefont {Perdew}\ \emph {et~al.}(1996)\citenamefont {Perdew},
  \citenamefont {Burke},\ and\ \citenamefont {Ernzerhof}}]{Perdew96}%
  \BibitemOpen
  \bibfield  {author} {\bibinfo {author} {\bibfnamefont {J.~P.}\ \bibnamefont
  {Perdew}}, \bibinfo {author} {\bibfnamefont {K.}~\bibnamefont {Burke}}, \
  and\ \bibinfo {author} {\bibfnamefont {M.}~\bibnamefont {Ernzerhof}},\ }\Doi
  {10.1103/PhysRevLett.77.3865} {\bibfield  {journal} {\bibinfo  {journal}
  {Phys. Rev. Lett.},\ }\textbf {\bibinfo {volume} {77}},\ \bibinfo {pages}
  {3865} (\bibinfo {year} {1996})}\BibitemShut {NoStop}%
\bibitem [{\citenamefont {Kresse}\ and\ \citenamefont
  {Hafner}(1993)}]{Kresse1993}%
  \BibitemOpen
  \bibfield  {author} {\bibinfo {author} {\bibfnamefont {G.}~\bibnamefont
  {Kresse}}\ and\ \bibinfo {author} {\bibfnamefont {J.}~\bibnamefont
  {Hafner}},\ }\Doi {10.1103/PhysRevB.47.558} {\bibfield  {journal} {\bibinfo
  {journal} {Phys. Rev. B},\ }\textbf {\bibinfo {volume} {47}},\ \bibinfo
  {pages} {558} (\bibinfo {year} {1993})}\BibitemShut {NoStop}%
\bibitem [{\citenamefont {Grimme}(2006)}]{Grimme2006}%
  \BibitemOpen
  \bibfield  {author} {\bibinfo {author} {\bibfnamefont {S.}~\bibnamefont
  {Grimme}},\ }\Doi {10.1002/jcc.20495} {\bibfield  {journal} {\bibinfo
  {journal} {Journal of Computational Chemistry},\ }\textbf {\bibinfo {volume}
  {27}},\ \bibinfo {pages} {1787} (\bibinfo {year} {2006})},\ ISSN \bibinfo
  {issn} {1096-987X}\BibitemShut {NoStop}%
\bibitem [{\citenamefont {Wei}\ \emph {et~al.}(2010)\citenamefont {Wei},
  \citenamefont {Jun-fang}, \citenamefont {Qinyu},\ and\ \citenamefont
  {Teng}}]{Wei2010}%
  \BibitemOpen
  \bibfield  {author} {\bibinfo {author} {\bibfnamefont {L.}~\bibnamefont
  {Wei}}, \bibinfo {author} {\bibfnamefont {C.}~\bibnamefont {Jun-fang}},
  \bibinfo {author} {\bibfnamefont {H.}~\bibnamefont {Qinyu}}, \ and\ \bibinfo
  {author} {\bibfnamefont {W.}~\bibnamefont {Teng}},\ }\Doi
  {10.1016/j.physb.2010.03.022} {\bibfield  {journal} {\bibinfo  {journal}
  {Physica B},\ }\textbf {\bibinfo {volume} {405}},\ \bibinfo {pages} {2498 }
  (\bibinfo {year} {2010})}\BibitemShut {NoStop}%
\bibitem [{\citenamefont {Tang}\ \emph {et~al.}(2009)\citenamefont {Tang},
  \citenamefont {Sanville},\ and\ \citenamefont {Henkelman}}]{Tang2009}%
  \BibitemOpen
  \bibfield  {author} {\bibinfo {author} {\bibfnamefont {W.}~\bibnamefont
  {Tang}}, \bibinfo {author} {\bibfnamefont {E.}~\bibnamefont {Sanville}}, \
  and\ \bibinfo {author} {\bibfnamefont {G.}~\bibnamefont {Henkelman}},\ }\href
  {http://stacks.iop.org/0953-8984/21/i=8/a=084204} {\bibfield  {journal}
  {\bibinfo  {journal} {J. Phys.: Condens. Matter},\ }\textbf {\bibinfo
  {volume} {21}},\ \bibinfo {pages} {084204} (\bibinfo {year}
  {2009})}\BibitemShut {NoStop}%
\bibitem [{\citenamefont {Sanville}\ \emph {et~al.}(2007)\citenamefont
  {Sanville}, \citenamefont {Kenny}, \citenamefont {Smith},\ and\ \citenamefont
  {Henkelman}}]{Sanville2007}%
  \BibitemOpen
  \bibfield  {author} {\bibinfo {author} {\bibfnamefont {E.}~\bibnamefont
  {Sanville}}, \bibinfo {author} {\bibfnamefont {S.~D.}\ \bibnamefont {Kenny}},
  \bibinfo {author} {\bibfnamefont {R.}~\bibnamefont {Smith}}, \ and\ \bibinfo
  {author} {\bibfnamefont {G.}~\bibnamefont {Henkelman}},\ }\Doi
  {10.1002/jcc.20575} {\bibfield  {journal} {\bibinfo  {journal} {J. Comput.
  Chem.},\ }\textbf {\bibinfo {volume} {28}},\ \bibinfo {pages} {899} (\bibinfo
  {year} {2007})}\BibitemShut {NoStop}%
\bibitem [{\citenamefont {Henkelman}\ \emph {et~al.}(2006)\citenamefont
  {Henkelman}, \citenamefont {Arnaldsson},\ and\ \citenamefont
  {J\'{o}nsson}}]{Henkelman2006}%
  \BibitemOpen
  \bibfield  {author} {\bibinfo {author} {\bibfnamefont {G.}~\bibnamefont
  {Henkelman}}, \bibinfo {author} {\bibfnamefont {A.}~\bibnamefont
  {Arnaldsson}}, \ and\ \bibinfo {author} {\bibfnamefont {H.}~\bibnamefont
  {J\'{o}nsson}},\ }\Doi {10.1016/j.commatsci.2005.04.010} {\bibfield
  {journal} {\bibinfo  {journal} {Comp. Mater. Sci.},\ }\textbf {\bibinfo
  {volume} {36}},\ \bibinfo {pages} {354 } (\bibinfo {year}
  {2006})}\BibitemShut {NoStop}%
\bibitem [{\citenamefont {Perdew}\ and\ \citenamefont
  {Zunger}(1981)}]{Perdew1981}%
  \BibitemOpen
  \bibfield  {author} {\bibinfo {author} {\bibfnamefont {J.~P.}\ \bibnamefont
  {Perdew}}\ and\ \bibinfo {author} {\bibfnamefont {A.}~\bibnamefont
  {Zunger}},\ }\Doi {10.1103/PhysRevB.23.5048} {\bibfield  {journal} {\bibinfo
  {journal} {Phys. Rev. B},\ }\textbf {\bibinfo {volume} {23}},\ \bibinfo
  {pages} {5048} (\bibinfo {year} {1981})}\BibitemShut {NoStop}%
\bibitem [{\citenamefont {Heyd}\ \emph {et~al.}(2003)\citenamefont {Heyd},
  \citenamefont {Scuseria},\ and\ \citenamefont {Ernzerhof}}]{heyd2003}%
  \BibitemOpen
  \bibfield  {author} {\bibinfo {author} {\bibfnamefont {J.}~\bibnamefont
  {Heyd}}, \bibinfo {author} {\bibfnamefont {G.~E.}\ \bibnamefont {Scuseria}},
  \ and\ \bibinfo {author} {\bibfnamefont {M.}~\bibnamefont {Ernzerhof}},\
  }\Doi {10.1063/1.1564060} {\bibfield  {journal} {\bibinfo  {journal} {J.
  Chem. Phys.},\ }\textbf {\bibinfo {volume} {118}},\ \bibinfo {pages} {8207}
  (\bibinfo {year} {2003})}\BibitemShut {NoStop}%
\bibitem [{\citenamefont {Heyd}\ \emph {et~al.}(2006)\citenamefont {Heyd},
  \citenamefont {Scuseria},\ and\ \citenamefont {Ernzerhof}}]{heyd2006}%
  \BibitemOpen
  \bibfield  {author} {\bibinfo {author} {\bibfnamefont {J.}~\bibnamefont
  {Heyd}}, \bibinfo {author} {\bibfnamefont {G.~E.}\ \bibnamefont {Scuseria}},
  \ and\ \bibinfo {author} {\bibfnamefont {M.}~\bibnamefont {Ernzerhof}},\
  }\Doi {10.1063/1.2204597} {\bibfield  {journal} {\bibinfo  {journal} {J.
  Chem. Phys.},\ }\textbf {\bibinfo {volume} {124}},\ \bibinfo {eid} {219906}
  (\bibinfo {year} {2006})}\BibitemShut {NoStop}%
\bibitem [{\citenamefont {Matsushita}\ \emph {et~al.}(2011)\citenamefont
  {Matsushita}, \citenamefont {Nakamura},\ and\ \citenamefont
  {Oshiyama}}]{Matsushita2011}%
  \BibitemOpen
  \bibfield  {author} {\bibinfo {author} {\bibfnamefont {Y.-i.}\ \bibnamefont
  {Matsushita}}, \bibinfo {author} {\bibfnamefont {K.}~\bibnamefont
  {Nakamura}}, \ and\ \bibinfo {author} {\bibfnamefont {A.}~\bibnamefont
  {Oshiyama}},\ }\Doi {10.1103/PhysRevB.84.075205} {\bibfield  {journal}
  {\bibinfo  {journal} {Phys. Rev. B},\ }\textbf {\bibinfo {volume} {84}},\
  \bibinfo {pages} {075205} (\bibinfo {year} {2011})}\BibitemShut {NoStop}%
\bibitem [{\citenamefont {Hedin}(1965)}]{Hedin1965}%
  \BibitemOpen
  \bibfield  {author} {\bibinfo {author} {\bibfnamefont {L.}~\bibnamefont
  {Hedin}},\ }\Doi {10.1103/PhysRev.139.A796} {\bibfield  {journal} {\bibinfo
  {journal} {Phys. Rev.},\ }\textbf {\bibinfo {volume} {139}},\ \bibinfo
  {pages} {A796} (\bibinfo {year} {1965})}\BibitemShut {NoStop}%
\end{thebibliography}
%

\end{document}